\documentclass[twocolumn]{aastex7}
\usepackage{threeparttable}
\usepackage{subfigure}
\usepackage{amsmath}
\usepackage{xspace}
\usepackage{booktabs}
\usepackage{lineno}
\usepackage{appendix}
\usepackage{pifont}

\newcommand{\cmtt}[1]{{\fontfamily{cmtt}\selectfont #1}}

\newcommand{\xmark}{\ding{55}}%

\newcommand{\ha}{H$\alpha$\xspace}
\newcommand{\hb}{H$\beta$\xspace}

\newcommand{\mgii}{\ion{Mg}{2}\xspace}

\newcommand{\nii}{[\ion{N}{2}]\xspace}

\newcommand{\oii}{[\ion{O}{2}]\xspace}
\newcommand{\oiii}{[\ion{O}{3}]\xspace}
\newcommand{\nev}{[\ion{Ne}{5}]\xspace}
\newcommand{\civ}{\ion{C}{4}\xspace}
\newcommand{\caii}{\ion{Ca}{2}\xspace}

\begin{document}
\shorttitle{Local Red Dots}
\shortauthors{Casey et al.}
\title{A Population of Little Red Dot-like Quasars in SDSS\footnote{Based on observations obtained with XMM-Newton, an ESA science mission with instruments and contributions directly funded by ESA Member States and NASA.}}

\correspondingauthor{Quinn O. Casey}
\author[0009-0005-9002-4800]{Quinn O. Casey}
\email[show]{quinn.o.casey.gr@dartmouth.edu}
\affil{Department of Physics and Astronomy, Dartmouth College, Hanover, NH 03755, USA}

\author[0000-0003-1468-9526]{Ryan C. Hickox}
\affiliation{Department of Physics and Astronomy, Dartmouth College, Hanover, NH 03755, USA}
\email{Ryan.C.Hickox@dartmouth.edu}

\author[0000-0001-7151-009X]{Nikko J. Cleri}
\affiliation{Department of Astronomy and Astrophysics, The Pennsylvania State University, University Park, PA 16802, USA}
\affiliation{Institute for Computational and Data Sciences, The Pennsylvania State University, University Park, PA 16802, USA}
\affiliation{Institute for Gravitation and the Cosmos, The Pennsylvania State University, University Park, PA 16802, USA}
\email{cleri@psu.edu}

\author[0000-0003-1420-6037]{Jonathan H. Cohn}
\affiliation{Department of Physics and Astronomy, Dartmouth College, Hanover, NH 03755, USA}
\email{Jonathan.Cohn@dartmouth.edu}

\author[0000-0002-5896-6313]{David M. Alexander}
\affiliation{Centre for Extragalactic Astronomy, Department of Physics, Durham University, South Road, Durham, DH1 3LE, UK}
\email{d.m.alexander@durham.ac.uk}

\author[0009-0004-9516-9593]{Emmanuel Durodola}
\affiliation{Department of Physics and Astronomy, Dartmouth College, Hanover, NH 03755, USA}
\email{emmanuel.a.durodola.gr@dartmouth.edu }

\author[0000-0002-8571-9801]{Kelly E. Whalen}
\affiliation{NASA Goddard Space Flight Center, Code 662, Greenbelt, MD 20771, USA}
\affiliation{Oak Ridge Associated Universities, NASA NPP Program, Oak Ridge, TN 37831, USA}
\email{kelly.e.whalen@nasa.gov}

\author[0000-0002-4684-9005]{Raphael E. Hviding}
\affiliation{Max-Planck-Institut für Astronomie, Königstuhl 17, D-69117 Heidelberg, Germany}
\email{hviding@mpia.de}

\author[0000-0001-8211-3807]{Tonima Tasnim Ananna}
\affiliation{Department of Physics and Astronomy, Wayne State University, Detroit, MI 48202, USA}
\email{tonima@wayne.edu}

\begin{abstract}
Compact and red sources in the high redshift ($z\sim5$) Universe, known as ``Little Red Dots" (LRDs), are among JWST's most intriguing discoveries. 
These sources have broad Balmer emission lines, weak X-ray emission, and unique spectral energy distributions (SEDs) poorly fit by either stellar or AGN templates.
Local LRD-like objects allow for detailed studies of the underlying physical processes with archival multi-wavelength datasets unavailable in the high-$z$ Universe. 
We show that the SDSS $ugriz$ filters at $z\approx0.4, 0.8$ overlap well with the JWST filters used to select LRDs at $z\sim5$. 
We use SDSS quasars to define a sample of $\sim1300$ Local Red Dots (LoRDs) which share the same photometric colors of LRDs.
A subset of the LoRD sample selected to have V-shaped continua ($N=244$) show prominent higher-order Balmer absorption features and \nev$\lambda$3426 emission, both of which would likely be missed in JWST/PRISM observations given the low spectral resolution. 
A composite SED of the LoRDs differs from a typical quasar SED in the rest-frame UV/optical, but the two agree with each other in the NIR.
The LoRD SED matches well with a stack of LRDs and can be modeled either as a reddened AGN combined with a host galaxy, or as a reddened AGN combined with a host galaxy and a cool blackbody.
Interestingly, the LoRDs are X-ray detected at a rate comparable to typical quasars. 
However, the probability that LoRDs and typical quasars would go undetected, if subject to the LRD X-ray upper limits, is $>50\%$.
\end{abstract}

\section{Introduction} \label{sec:intro}
The James Webb Space Telescope (JWST) has uncovered a puzzling population of sources at high redshift which are compact and red. 
These systems, frequently referred to as ``Little Red Dots" \citep[LRDs,][]{Matthee24}, have garnered immense attention due to their large number density in the early Universe \citep[e.g.,][]{Kokorev24, Kocevski25, Ma26}, weak (or no) X-ray emission \citep[e.g.,][]{Ananna24, Yue24, SachiBogdan25, Lambrides24}, and spectral energy distributions (SEDs) which push theoretical models to their limits \citep[e.g.,][]{Greene24, Kokorev24, Durodola25}. 

LRDs are characterized by a number of observables: broad Balmer emission lines, a ``V-shaped" UV-to-optical continuum, and a compact morphology \citep[e.g.,][]{Hviding25}. 
Our understanding of the optical spectral features primarily comes from JWST/PRISM observations \citep[e.g.,][]{Setton24, Hviding25}, but medium and high resolution grating observations are becoming more common in recent JWST cycles \citep[e.g.,][]{Kokorev25, Graaff26}. 
The broad ($\gtrsim1000 \text{ km s}^{-1}$) Balmer emission and red colors are often interpreted as originating from an active galactic nuclei (AGN; e.g., \citealt{Matthee24, Akins25, Inayoshi25b}); however, these observables can also be interpreted as stellar in origin \citep[e.g.,][]{Baggen24, Guia24, Kokubo24, Perez24}.
A significant population of LRDs exhibit narrow absorption features within their broad Balmer lines \citep[e.g.,][]{Inayoshi25, Matthee26, Davis26} which is highly unusual for typical QSOs.
The rest-frame continua of LRDs tend to be blue at wavelengths shorter than the Balmer limit (3645\AA) and red at the longer optical wavelengths, thus this change in slope forms a V-shape centered around the Balmer limit \citep[e.g.,][]{Setton24}. 
Numerous physical processes have been proposed to explain the Balmer breaks, such as
super-Eddington accretion \citep[e.g.,][]{Lambrides24, Liu25}, scattered UV light and a reddened QSO \citep[e.g.,][]{Akins25, Barro24, Greene24, Labbe25}, a gas-enshrouded black hole \citep[e.g.,][]{Inayoshi25, Naidu25}, among other scenarios \citep[e.g.,][]{Akins23, Labbe23, Williams24, Wang24, Wang25}.

The emission from LRDs has been constrained to varying levels of accuracy across the majority of the electromagnetic spectrum, and their SEDs are poorly fit with typical AGN and stellar templates.
The mid-infrared (MIR) slope of LRDs is relatively flat \citep[e.g.,][]{Williams24, Wang25}, which is inconsistent with observations of local AGN where the MIR rises due to emission from the dusty torus \citep[e.g.,][]{Hickox17}. 
The majority of individual LRDs have remained undetected in both far-infrared (FIR) and radio wavelengths \citep{Casey24, Williams24, Setton25-ALMA, Labbe25} which is also inconsistent with observations of local AGN. 
However, stacked observations suggest a non-negligible contribution from hot dust emission \citep{Delvecchio25, Perez26}. 
SED modeling has suggested the rest-frame UV emission originates from the host galaxy and the rest-frame optical is due to a reddened AGN \citep[e.g.,][]{Kokorev24, Kocevski25, Labbe25}, but conclusive evidence for the origin of these emission components has remained elusive. 

\begin{figure*}[ht]
    \centering
    \includegraphics[width=\linewidth]{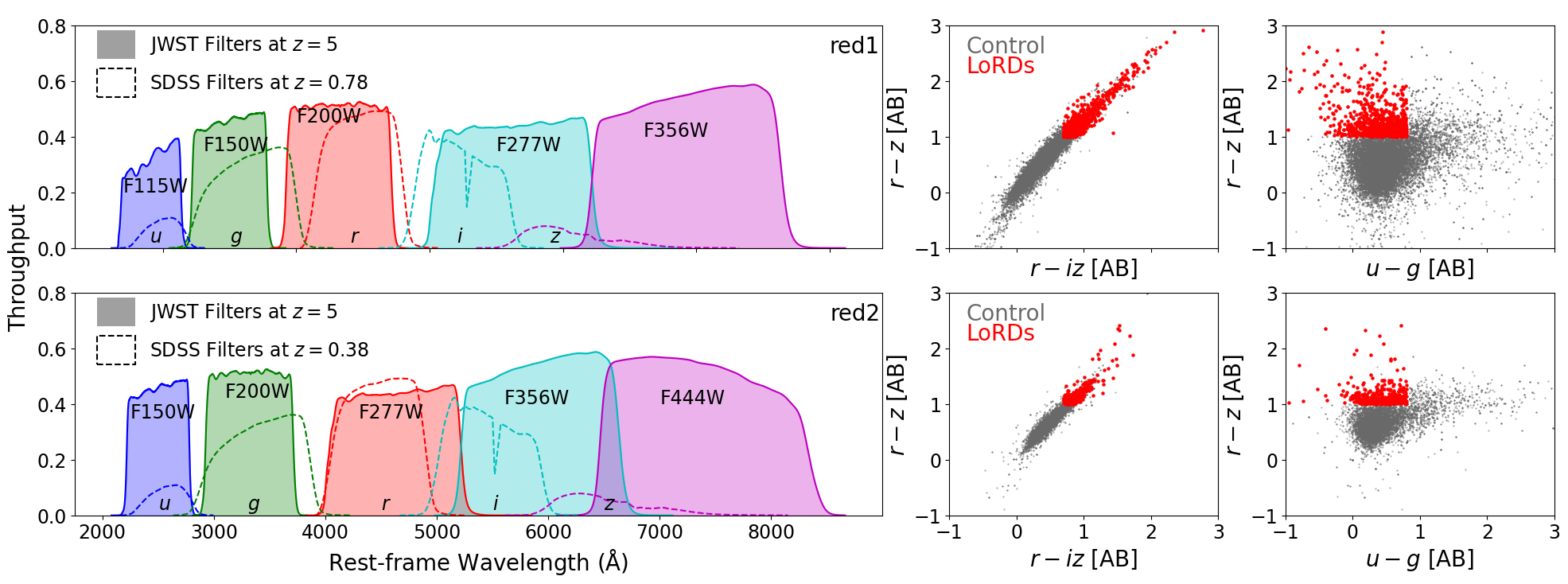}
    \caption{\textit{Left:} The throughput curves for the JWST filters at $z=5.0$ (shaded curves) compared to the SDSS filters (dotted lines) at $z=0.78$ (red1, top panel) and $z=0.38$ (red2, bottom panel). 
    The significant overlap between the filter sets allows us to select LoRDs using the SDSS filters. 
    \textit{Center:} $r-iz$ versus $r-z$ in AB magnitudes for the control sample (gray) and the LoRDs (red); the red1 sample is the top panel and red2 is the bottom panel.
    \textit{Right:} $u-g$ versus $r-z$ in AB magnitudes. }
    \label{fig:filters}
\end{figure*}

Given that many LRD interpretations require an AGN component, we expect these sources to be luminous ($L_X>10^{42} \text{ erg s}^{-1}$) in X-rays similar to nearby AGN \citep[e.g.,][]{Hickox18}. 
X-rays are created in AGN through the inverse Compton scattering of photons produced by the accretion disk, but LRDs are observationally weak in X-rays \citep[e.g.,][]{Ananna24, Furtak24, Yue24}. 
Results from stacking analyses suggest LRDs are $\sim10-100$ times weaker in X-rays than what is expected from local $L_{X}-L_{H\alpha}$ relations \citep[e.g.,][]{Ananna24, Yue24}.
Various explanations for the X-ray weakness include super-Eddington accretion \citep[e.g.,][]{Lambrides24, Pacucci24}, obscuration processes not observed in the local Universe \citep[e.g.,][]{Maiolino25, Naidu25, Lin25}, and scenarios without an AGN \citep[e.g.,][]{Akins25, Baggen24, Williams24}. 

The number density of LRDs is $\sim10-100$ times more numerous than UV-selected quasars (QSOs) of similar magnitudes between $4\lesssim z \lesssim 8$ \citep[e.g.,][]{Kokorev24, Kocevski25}. 
At $z\lesssim4$ the number density of LRDs has been shown to decline rapidly \citep[e.g.,][]{Kocevski25, Ma26} which may be related to the ``inside-out" growth of galaxies \citep[e.g.,][]{vandokkum14, Kocevski25}.
Due to this decreasing number density, as well as a lack of rest-frame UV spectroscopic observations, much of our understanding of these sources comes from high redshift observations which are observationally expensive. 
\citet{Lin25} discovered a handful of nearby LRDs ($z=0.1-0.2$) in the Sloan Digital Sky Survey (SDSS) by projecting LRD spectra into \textit{Galaxy Evolution Explorer} (\textit{GALEX}; \citealt{Martin05}) and SDSS photometry to determine what their local colors would look like. 
One of these sources was also independently discovered by \citet{Ji25b} by searching for V-shaped spectra in a catalog of broad-line dwarf galaxies. 
Likewise, \citet{Lin26b} identified 27 LRDs at $z=0.2-0.9$ by selecting broad line sources in the Dark Energy Spectroscopic Instrument (DESI; \citealt{DESI26}) DR1 catalog which share similar emission line properties to the LRDs at high-$z$.
These local analog candidates provide higher quality observations which can be used to test theoretical models, but the small number statistics make it difficult to say anything on a population scale. 

In this paper we take advantage of both the redshift range ($z\sim 0.4, 0.8$) where the rest-frame wavelengths probed by SDSS and certain JWST filters (at $z\sim5$) overlap with each other, as well as the wealth of archival data at these redshifts. 
Compared to previous local LRD studies, our sample is much larger, due to our rest-frame matched selection, which allows us to treat the LRD-like sources in a statistical sense.
Section \ref{sec:data} describes the parent sample and how we select LRD-like objects. 
We investigate the multiwavelength properties of the local red dots (LoRDs) in Section \ref{sec:mw-analysis}. 
In Section \ref{sec:disc-conc} we discuss the impact of morphology, spectral resolution, and X-ray upper limits on our understanding of the similarity between LoRDs and LRDs.
We summarize our findings and conclude with potential use cases of the LoRDs in Section \ref{sec:conclusions}.
When necessary we utilize the following cosmology: $H_{0}=70\text{ km s}^{-1} \text{ Mpc}^{-1}$, $\Omega_{M}=0.3$, and $\Omega_{\Lambda}=0.7$ \citep{Tegmark04}.

\section{Data and Sample Selection} \label{sec:data}

We draw from the SDSS DR16 quasar (DR16Q; \citealt{Lyke20}) catalog of \citet{Wu22} which contains optical spectroscopy for $\sim 750,000$ broad line quasars which were observed during the SDSS-I/II/III/IV campaigns \citep[][]{York00, Eisenstein11, Blanton17}.
The SDSS-I/II legacy spectrographs were used through DR7; these observations use 3\arcsec\ diameter fibers and cover a wavelength range of $\sim3800-9200$\AA. 
SDSS-III/IV use the upgraded Baryon Oscillation Spectroscopic Survey (BOSS) spectrographs \citep{Smee13} which employ 2.5\arcsec\ diameter fibers and have a wavelength coverage of $\sim3650-10400$\AA.
Both spectrographs have a spectral resolution of $R\sim2000$.
If a source was observed in multiple campaigns then the SDSS-IV observation is adopted as the primary spectrum in the DR16Q catalog.

The DR16Q spectra are fit with a global continuum+emission line model using the \cmtt{PyQSOFIT} code described in \citet{Guo18}, \citet{Shen16,Shen19}, and \citet{Wu22}.
Along with emission line measurements, the DR16Q catalog contains single-epoch virial black hole (BH) masses, bolometric luminosities (assuming a typical AGN), and updated systemic redshifts. 
As described in \citet{Wu22}, bolometric luminosities are estimated from continuum luminosities at rest-frame wavelengths of 1350, 3000, or 5100\AA, depending on the redshift of the source (3000\AA\ is the default).
BH masses are estimated from \hb ($z<0.7$), \mgii ($0.7\leq z\lesssim2.0$), and \civ ($z\gtrsim 2.0$) where the redshifts change slightly based on the spectrograph used. 
This quasar sample covers a broad range of redshifts ($0.1\lesssim z \lesssim 6.0$) and luminosities ($44 \lesssim \log(L_{\text{bol}}/\text{erg s}^{-1})\lesssim48$). 

\subsection{Photometric Selection} \label{subsec:selection}
\citet{Greene24} introduced two photometric criteria to select red JWST sources, the majority of which are LRDs: red1 = (F115W - F150W $<$ 0.8) $\wedge$ (F200W - F277W $>$ 0.7) $\wedge$ (F200W - F356W $>$ 1.0), and red2 = (F150W - F200W $<$ 0.8) $\wedge$ (F277W - F356W $>$ 0.7) $\wedge$ (F277W - F444W $>$ 1.0). 
They also require a compactness criterion which ensures the majority of the flux is dominated by a point source. 
The redshift distribution of LRDs peaks around $z\approx5$ \citep[e.g.,][]{Kocevski25} which means the rest-frame red1 and red2 selection criteria probe the near-UV and optical wavelength ranges. 
There exists a large amount of archival data which cover this wavelength range in the local Universe, and we use these data to search for LRD-like objects.

To select local LRD-like objects (i.e., LoRDs) in SDSS we identify redshift bins where the JWST and SDSS filters have significant rest-frame overlap. 
We do this by minimizing a mismatch function, $M(z)$ (Equation \ref{eq:mismatch}) which evaluates the overlap between the filters used in the red1/red2 selection and the SDSS $ugriz$ filter set. 
\begin{equation}\label{eq:mismatch}
\begin{split}
    M(z) = \sum_i \biggl[{} & \biggl( \frac{\lambda_{\text{blue}}^{\mathrm{JWST}, i}}{1+5} - \frac{\lambda_{\text{blue}}^{\mathrm{SDSS}, i}}{1+z} \biggr)^2 \\
    & + \biggl( \frac{\lambda_{\text{red}}^{\mathrm{JWST}, i}}{1+5} - \frac{\lambda_{\text{red}}^{\mathrm{SDSS}, i}}{1+z} \biggr)^2 \biggr]
\end{split}
\end{equation}
In Equation \ref{eq:mismatch}, $i$ represents the filters used in the red1/red2 selection, and $\lambda_{\text{blue}}$ and $\lambda_{\text{red}}$ are the blue and red edges of the filter, respectively.
We find the red1 and red2 filters at $z=5.0$ are offset the least when the SDSS filters are redshifted to $z=0.78$ and $z=0.38$, respectively (see Figure \ref{fig:filters}). 
In the rest-frame, the SDSS $i$ and $z$ bands both overlap with the F277W (red1 selection) and the F356W (red2 selection) filters. 
To remedy this we define an ``$iz$" band which is the logarithmic average of the fluxes in those two bands. 
Therefore, at $z=0.78$, we can map the JWST filters to the SDSS filters as follows: F115W maps to $u$, F150W to $g$, F200W to $r$, F277W to $iz$, and F356W to $z$.
Similarly, at $z=0.38$, the filter F150W maps to $u$, F200W to $g$, F277W to $r$, F356W to $iz$, and F444W to $z$.
Given this mapping, we can apply the red1/red2 selection criteria using the SDSS filters as shown in Equations \ref{eq:red1} and \ref{eq:red2} (colors in AB magnitudes):
\begin{multline}\label{eq:red1}
    \text{local red1} = (u-g<0.8) \wedge (r-iz>0.7) \wedge \\ (r-z>1.0) \wedge (0.68<z_{\text{spec.}}<0.88)
\end{multline}
\begin{multline}\label{eq:red2}
    \text{local red2} = (u-g<0.8) \wedge (r-iz>0.7) \wedge \\ (r-z>1.0) \wedge (0.28<z_{\text{spec.}}<0.48).
\end{multline}

\begin{figure}
    \centering
    \includegraphics[width=\linewidth]{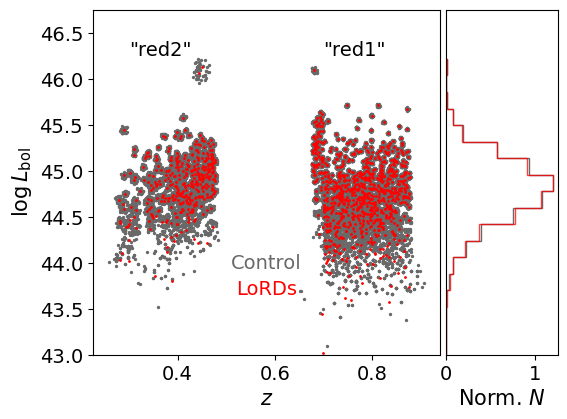}
    \caption{Distribution of LoRDs and control QSOs in redshift-bolometric luminosity space. 
    LoRDs are selected based on Equations \ref{eq:red1} and \ref{eq:red2}, and control QSOs are selected to match LoRDs in redshift-bolometric luminosity space allowing for double counting. }
    \label{fig:redshift-luminosity}
\end{figure}

We select quasars from DR16Q which pass the local red1/red2 selection criteria and have redshifts within $z\pm0.1$ of the optimized redshift values. 
We also require sources to be brighter than 22.2 magnitudes in the $i$-band (i.e., the median $5\sigma$ depth in SDSS; \citealt{Dawson16}).
The red1 and red2 selections yield 912 and 383 LoRDs which corresponds to $\approx 1.8\%$ and $\approx 6.8\%$ of the entire DR16Q sample in their respective redshift ranges.
We also define a QSO ``control" sample to match the LoRD sample in redshift and bolometric luminosity. 
Figure \ref{fig:filters} (center and right panels) shows our selection of LoRDs and the control sample in color-color space.
To select the control sample, we define a rectangle around each LoRD with widths 0.05 (in $\Delta z$) and 0.3 dex (in bolometric luminosity) and select the nearest twenty five QSOs within the rectangle for each LoRD, allowing for multiple counts of the same source to ensure a similar luminosity distribution between the LoRDs and the control sample (see Figure \ref{fig:redshift-luminosity}).
There are $\sim23,000$ and $\sim9,600$ control QSOs in the red1 and red2 samples in total, but only $\sim14,000$ and $\sim5,500$ unique sources in each redshift bin.

SDSS required an unresolved optical morphology in order for a source to be targeted as a quasar \citep{Ross12, Myers15}, but the \citet{Wu22} QSO catalog is not necessarily built from SDSS targeting, and therefore some sources are resolved in the optical. 
We do not impose any explicit compactness criteria in our selection given that the angular resolution of SDSS and JWST differ by more than an order of magnitude (e.g., $r_{\text{SDSS, phys}}\approx6.4 \text{ kpc}$, $r_{\text{JWST, phys}}\approx0.4 \text{ kpc}$ at $z=0.4$). 
Furthermore, whether or not the physical processes which produce LRD-like emission can exist within a normal sized (e.g., Milky Way like) host galaxy at low-$z$ is an open question.
Of the 1295 LoRDs, 59\% are classified as stars (i.e., point source like) by the SDSS imaging pipeline \citep{Lupton01}, compared to 92\% of the control sample. 
In Section \ref{subsec:morphology} we discuss the impact of compactness on our sample and how this alters our interpretation of a subset of sources.

\section{Multiwavelength Analysis} \label{sec:mw-analysis}


\subsection{Optical Spectroscopy} \label{sec:spec}

\begin{table*}[]
    \caption{Sample of photometrically selected LoRDs. $z$ and $\log L_{\rm{bol}}$ are from the DR16Q catalog \citep{Wu22}, the V-shape criteria are described in Section \ref{sec:spec}, and we determine $\log L_{X,\text{ 2-10 keV}}$ in Section \ref{sec:x-ray}.}
    \begin{tabular}{ccccccc}
    \hline \hline
Name & $z$ & $\log L_{\rm{bol}}$ & V-Shape & $\beta_{\rm{UV}}$ & $\beta_{\rm{opt}}$ & $\log L_{X,\text{ 2-10 keV}}$ \\ 
 &  & [erg s$^{-1}$] &  &  &  & [erg s$^{-1}$] \\ 
\hline
SDSS J231224.75+290710.9 & 0.28 & $44.7\pm0.01$ & False & $-0.27\pm$0.14 & $0.01\pm$0.02 & -- \\
SDSS J132700.12+490344.4 & 0.28 & $44.4\pm0.01$ & False & $-0.12\pm$0.07 & $0.01\pm$0.02 & -- \\
SDSS J092621.14+310847.8 & 0.28 & $44.5\pm0.0$ & True & $0.01\pm$0.04 & $0.54\pm$0.01 & 42.8 $\pm$ 0.1 \\
SDSS J104947.35+452323.2 & 0.28 & $44.0\pm0.02$ & -- & $-0.15\pm$0.07 & $0.75\pm$0.04 & -- \\
SDSS J010906.64+001825.2 & 0.28 & $44.1\pm0.01$ & -- & $-0.02\pm$0.05 & $0.76\pm$0.02 & -- \\
    \hline
    \end{tabular}
    \label{tab:sample}
    \begin{center}
        (This table is available in its entirety in a machine-readable form in the online journal.) 
    \end{center} 
\end{table*}

\begin{figure*}
    \centering
    \includegraphics[width=\linewidth]{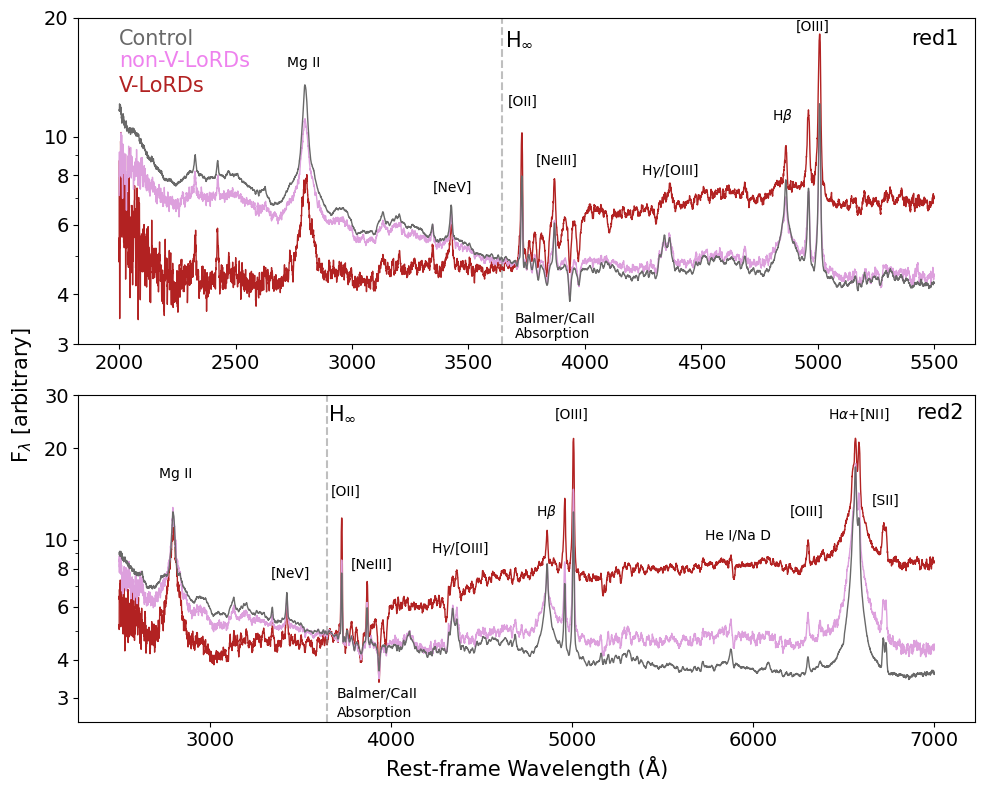}
    \caption{Median stacked spectra for the red1/red2 LoRDs as well as the control sample. 
    The spectra are normalized to the integrated flux between $3645\pm100$\AA\ and scaled by a factor of 1000. 
    The red spectra are selected to have V-shaped continua (top $N=111$, bottom $N=133$) and the pink spectra do not (top $N=229$, bottom $N=143$). 
    Prominent emission/absorption features and the Balmer limit ($H_{\infty}$) are labeled.
    The stacked spectra for the control sample is shown in gray. }
    \label{fig:stacked_spec}
\end{figure*}

Many LRDs are selected from photometry alone, but numerous studies have shown that sources which do not share the same physical properties of LRDs may contaminate this selection \citep[e.g.,][]{Hviding25, Labbe25} such as brown dwarfs \citep{Langeroodi23}. 
Therefore, selecting LRDs with high confidence usually requires spectroscopic determinations of redshift, broad emission lines, and a V-shaped continuum. 
The spectroscopic identification of a non-zero redshift and broad emission lines can easily distinguish LRD SEDs from the SEDs of brown dwarfs.
Interestingly, \citet{Hviding25} showed that in the high-$z$ Universe nearly all ($\sim80\%$) point source objects with V-shaped spectral continua also have broad Balmer lines. 
Each object in the DR16Q catalog is a broad-line source, so we need not trim our LoRD/control samples on this criterion.

We adopt an approach similar to \citet{Setton24} and \citet{Hviding25} to determine whether the LoRDs have V-shaped spectral continua. 
We first mask prominent emission lines (i.e., \mgii$\lambda\lambda 2797, 2803$, \oii$\lambda 3727$, \hb, \oiii$\lambda \lambda 4959, 5007$, \ha, \nii$\lambda 6584$), and require a median signal-to-noise ratio greater than three per pixel on the spectra between 3000 and 5000\AA\xspace to eliminate noisy spectra.
We fit the SDSS spectra with a broken power law function of the form $f_{\lambda}=a\lambda_{\text{rest}} ^{\beta}$, and allow the break to occur anywhere between $\lambda_{\text{rest}}=3645\pm100\text{\AA}$. 
We fit the broken power law to the spectra using \cmtt{emcee} \citep{Foreman13} and consider a photometrically selected LoRD to be ``V-shaped" if the following criteria are true: $\beta_{\text{opt}} > 0.5$ at the $1\sigma$ level and $\beta_{\text{opt}} > 2\beta_{\text{UV}}$.
These criteria differ from the selection of LRDs at high-$z$ (e.g., \citealt{Hviding25} requires $\beta_{\text{UV}}<-0.2$, $\beta_{\text{opt}} > 0$, and $\beta_{\text{opt}}-\beta_{\text{UV}} > 0.5$) given that growing evidence suggests the rest-UV component of LRDs has a substantial stellar contribution \citep[e.g.,][]{Cloonan26, Perez26} and the stellar populations of LRDs and LoRDs may differ.
In the red1 and red2 samples we classify 111 and 133 LoRDs as V-shaped, respectively (henceforth referred to as V-LoRDs). 
In red1 (red2) there are 229 (143) sources without a V-shaped continuum (i.e., non-V-LoRDs) and 572 (107) sources which are too noisy to fit.
The sample of LoRDs is presented in Table \ref{tab:sample}.

\begin{table*}[t]
\caption{Comparison of spectroscopic features between the high-$z$ LRDs, V-LoRDs, non-V-LoRDs, and the control QSOs. }
    \centering
    \begin{tabular}{c|c|c|c|c}
    \hline \hline
        Spectroscopic Feature & LRDs & V-LoRDs & non-V-LoRDs & Control QSOs \\
        \hline
        UV Continuum & Blue & Blue/Flat & Blue  & Blue \\
        Optical Continuum & Red & Red & Blue/Flat  & Blue/Flat \\
        Broad \mgii Emission & \checkmark$^{a}$ & \checkmark & \checkmark & \checkmark \\
        \nev Emission & \xmark$^{b}$ & \checkmark & \checkmark & \checkmark \\
        H$\epsilon$-H12 Absorption & ? & \checkmark & \checkmark & \checkmark \\
        \caii H \& K Absorption & ? & \checkmark & \checkmark & \checkmark \\
        Broad \ha/\hb & \checkmark & \checkmark & \checkmark & \checkmark \\
        Narrow \ha/\hb Absorption$^{c}$ & $\sim40-60\%$ & ? & ? & $\sim0\%$ \\
        \hline
        \end{tabular}
        \centering
        \begin{tablenotes}
            \item $^{a}$ \mgii is frequently undetected in individual LRDs, but it is detected in stacks of LRD spectra \citep{Perez26}. 
            \item  $^{b}$ \nev$\lambda$3426 has only been detected in one LRD \citep{Labbe24}; however, identifying this line in the low-resolution PRISM data may be difficult (see Section \ref{subsec:spec}). 
            \item  $^{c}$ Estimates of narrow Balmer absorption in LRDs come from \citet{Davis26} and \citet{Matthee26}. We do not quantify Balmer absorption in our sample (see Section \ref{sec:spec}), but show examples of potential absorption in the Appendix. We assume the control QSOs have $\sim0\%$ absorption based on the results from \citet{Shangguan26}. 
        \end{tablenotes}
    \label{tab:spec-compare}
\end{table*}

Our V-shape criteria could, in principle, select sources with a pure rising red continuum and a weak break around the Balmer limit which would serve as a poor candidate LRD analog. 
However, we choose to retain sources with modestly positive UV slopes relative to their optical slopes in our V-LoRD sample as many of these sources have the most positive optical slopes. 
For V-LoRDs with a UV slope greater than zero, the mean UV slope is $\langle\beta_{\rm{UV}}\rangle=0.19\pm0.15$ compared to $\langle\beta_{\rm{opt}}\rangle=1.3\pm0.51$ in the optical, where the quoted uncertainty is the standard deviation.
The majority of our analysis focuses on the entire LoRD sample, so this should not bias our results, but we discuss caveats in Section \ref{sec:disc-conc}.

Figure \ref{fig:stacked_spec} shows the median stacked spectra for the V-LoRDs and non-V-LoRDs in red and pink, respectively; control QSOs are shown in gray. 
To perform the stack we shift each spectrum into its rest-frame (using the systemic redshifts from DR16Q) and smooth it using SciPy's \cmtt{medfilt} function with a kernel size of seven. 
We interpolate the smoothed rest-frame spectra along a wavelength grid spanning $\lambda=2000-5500\text{\AA}$ for the red1 sample and $\lambda=2500-7000\text{\AA}$ for the red2 sample; in both cases we use wavelength bins of 0.5\AA. 
We require each flux measurement and the associated error to be greater than 0 and normalize each spectrum to the integrated flux between $3645\pm100$\AA.

There exist clear differences between the V-LoRDs and the non-V-LoRDs/control sample in both redshift bins. 
The control and non-V-LoRDs have declining UV slopes, whereas the V-LoRDs are relatively flat. 
The stacked spectra of the control sample and the non-V-LoRDs show a similar blue rest-optical continuum, which is in stark contrast to the V-LoRDs' red rest-optical.
None of the stacked spectra show a steep jump at the Balmer limit which is seen in some of the high-$z$ LRDs \citep[c.f., ``The Cliff,"][]{Graaff25}.
Some individual spectra ($\sim$ a few percent) do exhibit this steep Balmer break regardless of their V-shaped classification, but it is not ubiquitous in the LoRD spectra. 
There are significant Balmer (H$\epsilon$-H12) and \caii (H and K) absorption features in both V- and non-V-LoRDs relative to the control sample ($3750\rm{\AA} \lesssim \lambda \lesssim 3970\rm{\AA}$); these features are more prevalent in the red1 sample.
The LoRDs and the control sample both show prominent emission at \nev$\lambda3426$. 
$\rm{Ne}^{4+}$ requires 97 eV; thus, \nev emission is most often attributed to luminous AGN \citep[e.g.,][]{Gilli10, Cleri23a, Cleri23b, Negus23}.
We compare the features in these stacked spectra to LRDs in Table \ref{tab:spec-compare}, and interpret the LoRD spectra in the context of LRDs in Sections \ref{subsec:morphology} and \ref{subsec:spec}.

A unique feature in high-$z$ LRD spectra is narrow absorption within the broad Balmer emission lines. 
The exact percentage of LRDs with this absorption feature is uncertain as its prominence increases with increased spectral resolution/sensitivity, and the number of LRDs observed with medium/high-resolution spectroscopy is currently meager.
Current estimates suggest $\sim 40-60\%$ of LRDs \citep[e.g.,][]{Davis26, Matthee26} have this feature, and it is more commonly seen in the \ha emission line compared to \hb. 
We manually inspect the LoRD \ha and \hb emission lines and find that a substantial number of sources show signatures of narrow absorption within their broad lines. 
However, trying to prove this Balmer absorption in the SDSS spectra would be challenging as low signal-to-noise and sky lines complicate the identification of this feature. 
Therefore, we choose not to quantify the Balmer absorption in this work as new observations would likely be necessary. 
We show a select number of LoRD spectra in the Appendix which display signatures of this absorption feature; fitting the feature is beyond the scope of this work.

\subsection{UV/Optical/NIR SEDs} \label{subsec:seds}
Standard SED modeling (e.g., \cmtt{CIGALE}/\cmtt{PROSPECTOR}) of LRDs has consistently struggled to fully reproduce the observations \citep[e.g.,][]{Williams24, Ronayne25}. 
To understand the differences between the control QSOs and LoRDs, we construct composite SEDs with archival SDSS and the \textit{Wide-field Infrared Survey Explorer} (WISE; \citealt{Wright10}) photometry using the same approach as \citet{Hickox17}. 
For each source in the control and LoRD samples we require $\text{S/N} > 3$ in each band. 
We exclude the less sensitive W4 band from the composite SEDs to retain more sources in our sample. 
In total we have $N\sim400$ LoRDs and $N\sim10000$ control QSOs which contribute to the composite SED, around $\sim30\%$ of each sample size. 
If we included W4 we would only retain $\sim10\%$ of either sample.

\begin{figure}
    \centering
    \includegraphics[width=\linewidth]{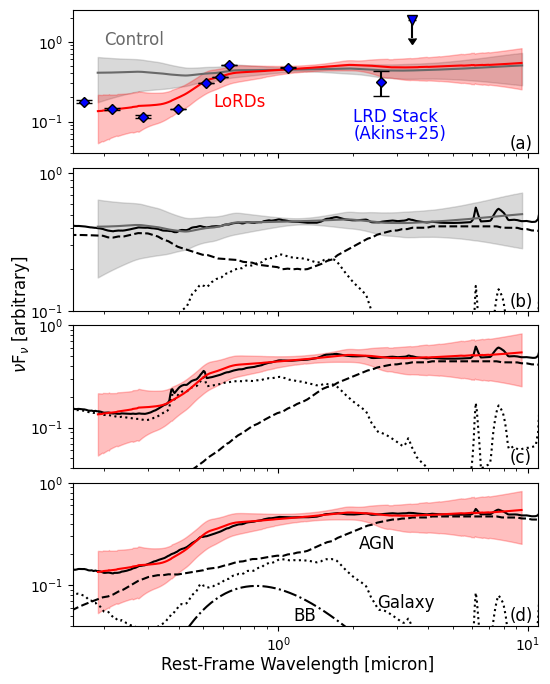}
    \caption{Composite SEDs for the control (gray) and LoRD (red) samples; in panel (a) we show a stack of LRDs from \citet{Akins25} in blue. 
    The solid gray/red lines represent the mean SED and the shaded region is one standard deviation from the mean. 
    In panels (b), (c), and (d) we show fits to the composite SEDs in black. 
    Solid black shows the cumulative fit, dashed lines represent the AGN component, dotted lines are the galaxy component, and the dot-dashed line is the blackbody feature.
    Panels (c) and (d) illustrate that the LoRD SED can be fit well with or without the blackbody component.}
    \label{fig:seds}
\end{figure}

To build the SEDs we shift each flux measurement into its rest-frame and interpolate between the fluxes (in $\nu F_{\nu}$) using piecewise power laws in log-log space.
We normalize each individual SED by the integrated flux between 0.4 and 4 microns which roughly traces the stellar component in typical AGN. 
We remove individual flux measurements more than $6\sigma$ from the mean and average the flux at each wavelength bin. 
The composite SEDs are shown in Figure \ref{fig:seds}. 
For comparison purposes we include a stack of LRDs from \citet{Akins25} over the same wavelength range as the composite SEDs; the filters contributing to the LRD stack that is shown are JWST/F115W, F150W, F200W, F277W, F356W, F410W, F444W, F770W, F1800W, and Spitzer/MIPS 24$\mu$m (upper-limit). 

As a qualitative exercise to better understand the sample differences, we perform simple multi-component fits to the composite SEDs. 
The components we include are a type 1 AGN continuum template from \citet{Richards06}, galaxy templates from \citet{Assef10} (elliptical, spiral, and irregular components), the empirical emission line galaxy (ELG) SED from \citet{Forrest18}, and a cool ($T_{\rm{BB}}\sim4000$K) blackbody. 
We allow the AGN and galaxy templates to be dust reddened according to the prescription of \citet{Calzetti00}. 
A cool blackbody is frequently invoked when modeling LRD spectral continua and is interpreted as an optically thick gas envelope surrounding the supermassive black hole (SMBH; e.g., \citealt[]{Graaff26}); however, we remain agnostic to what this feature physically represents, and discuss the implications of the blackbody component in Section \ref{subsubsec:physical-models}.
To fit the SEDs, we adjust the components by hand until we find a combination of parameters (i.e., normalization, $A_V$, $T_{\rm{BB}}$) which match the composite SEDs well. 
We optimize the normalization by fitting the templates to the composite SEDs using SciPy's \cmtt{nnls} function. 

The control sample is well fit with a minimally reddened AGN which dominates the flux, as well as a mixture of elliptical/spiral components.
This is similar to the result of \citet{Hickox17} with slightly different contributions.
We show two different fits to the LoRD SED in Figure \ref{fig:seds} (panels c and d) to demonstrate that the SEDs can be fit well with or without a cool blackbody component.
Both fits require a dominant reddened AGN ($1.0\lesssim A_V \lesssim 2.0$ mag) which is needed to match the NIR, a slightly reddened elliptical component in the optical, and a small contribution from the ELG in order to match the UV. 
The differences between the fits are a function of AGN reddening as well as the mechanism which produces the increased flux at the Balmer limit: a cool blackbody or the irregular (i.e., starburst) galaxy template. 

We note the galaxy component in Figure \ref{fig:seds} (d) is sub-dominant despite the presence of \caii H \& K absorption features in the stacked spectra (Figure \ref{fig:stacked_spec}). 
However, the relative strength of the galaxy component at $\lambda\approx4000$\AA\xspace ($\sim30\%$) should produce such absorption features with non-zero EWs. 
Adopting typical EW values for the various galaxy components ($\sim4-17$\AA; \citealt{Bica88, Fritz14}), the \caii K EW should be $\sim2.7$\AA\xspace (\caii H is blended with H$\epsilon$). 
We stack the spectra of the LoRDs used to create the composite SED shown in Figure \ref{fig:seds} (d) and fit the \caii K absorption line. 
The measured EW is $\sim3.4$\AA\xspace which is consistent with the galaxy components used to fit the SED. 


\begin{figure}
    \centering
    \includegraphics[width=\linewidth]{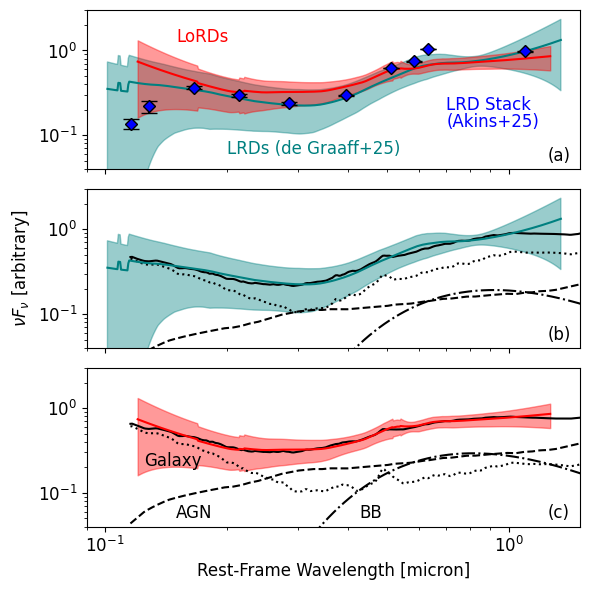}
    \caption{Composite SEDs for the LRDs (teal) and the LoRDs (red); the LRD stack from \citet{Akins25} is shown as blue diamonds. 
    Similar to Figure \ref{fig:seds}, panels (b) and (c) show exploratory fits to the SEDs in black, where the solid line is the cumulative fit, the dotted line is the galaxy component, dashed line is the AGN component, and the dot-dashed line is the blackbody feature.}
    \label{fig:seds-lrds}
\end{figure}

In order to more directly compare the SEDs of LoRDs and LRDs, we create a composite SED using the LRDs from \citet{Graaff26}. 
The \citet{Graaff26} LRD sample ($N=116$) was selected spectroscopically and thus likely suffers from less contamination than the sources in the \cite{Akins25} stack. 
We include the following filters in the LRD stack: HST/F814W, JWST/F090W, F115W, F150W, F200W, F277W, F356W, and F444W.
To create the LoRD SED and compare across a similar rest-frame wavelength range we use \textit{GALEX}/NUV, SDSS/$u$, $g$, $r$, $i$, $z$, and the UKIRT Infrared Deep Sky Survey (UKIDSS, \citealt{Lawrence07}) $Y$, $J$, and $H$ bands.
We require $3\sigma$ detections in each band and are left with 35 LRDs and 71 LoRDs. 
We construct the composite SEDs in the same fashion as previously described, except we normalize the individual SEDs between 0.2 and 1.3 microns; the SEDs are shown in Figure \ref{fig:seds-lrds}.
Both SEDs are fit with slightly different combinations of galaxy, blackbody, and AGN components. 
In both cases the UV is fit with the ELG template, and the Balmer break is produced by a combination of the elliptical template, a cool blackbody ($T_{\text{BB, LoRD}}=4500$K, $T_{\text{BB, LRD}}=4000$K), and a reddened AGN ($A_V=1.0$). 
We emphasize that the fits in Figures \ref{fig:seds} and \ref{fig:seds-lrds} are merely exploratory and not meant to be interpreted as definitive.
The composite SEDs are presented in Table \ref{tab:seds}.

\begin{table}[]
    \caption{The composite SEDs presented in Section \ref{subsec:seds} and Figures \ref{fig:seds} and \ref{fig:seds-lrds}. The flux measurement is the mean and the quoted uncertainty is one standard deviation from the mean; the units on $\nu F_{\nu}$ are arbitrary due to the normalization. The SEDs cover different wavelength ranges, therefore some short and long wavelengths may not have a corresponding flux measurement.}
    \centering
    \begin{tabular}{ccccc}
    \hline \hline
         & Control$^{a}$ & LoRD$^{b}$ & LRD$^{c}$ & LoRD$^{d}$ \\
       $\lambda \text{ }(\mu\rm{m})$ & $\nu F_{\nu}$  & $\nu F_{\nu}$ & $\nu F_{\nu}$ & $\nu F_{\nu}$ \\
       \hline 
        0.1013 & -- & -- & 0.352 $\pm$ 0.379 & -- \\
        0.1018 & -- & -- & 0.351 $\pm$ 0.376 & -- \\
        0.1023 & -- & -- & 0.349 $\pm$ 0.373 & -- \\
        0.1029 & -- & -- & 0.348 $\pm$ 0.370 & -- \\
        0.1034 & -- & -- & 0.347 $\pm$ 0.367 & -- \\
    \hline
    \end{tabular}
    \label{tab:seds}
    \\
    (This table is available in its entirety in a machine-readable form in the online journal.)
    \begin{tablenotes}
        \item $^{a}$ The control SED as shown in Figure \ref{fig:seds} (gray). 
        \item $^{b}$ The LoRD SED as shown in Figure \ref{fig:seds} (red). 
        \item $^{c}$ The SED using the LRDs from \citet{Graaff26}, as shown in Figure \ref{fig:seds-lrds} (teal).
        \item $^{d}$ The LoRD SED as shown in Figure \ref{fig:seds-lrds} (red). 
    \end{tablenotes}
\end{table}

\subsection{X-Ray Properties} \label{sec:x-ray}

Given the X-ray weakness observed in the LRDs, we seek to determine whether the LoRDs are X-ray weak as well. 
We cross-match the LoRDs and the control sample to the XMM-Newton Serendipitous Source Catalog DR14 (4XMM-DR14; \citealt{Webb20}) 
using a maximum matching radius of 4\arcmin. 
We consider X-ray sources within 10\arcsec\ of the SDSS source to be a true match and require $\text{S/N}>3$ to count the source as a detection. 
Above the 10\arcsec\ threshold, matches are dominated by random (chance) alignments rather than true associations; 2166 control QSOs and 143 LoRDs are covered by archival XMM observations. 
The detection fractions in the soft ($0.5-2$ keV), full ($0.5-10$ keV), and hard ($2-10$ keV) bands are reasonably consistent with each other across the control and LoRD samples (see Table \ref{tab:x-ray-properties}). 
For sources which were covered with XMM but not detected, we estimate the flux upper limits using the HIgh-energy LIght curve GeneraTor (HILIGT; \citealt{Saxton22}) assuming a spectral model with a power law slope of $\Gamma=2.0$ \citep[e.g.,][]{Nandra94}.


Given the LoRDs largely appear to have normal X-ray detection fractions compared to the control sample (i.e., $<2\sigma$ difference in each band), we compare their X-ray luminosities to the stacking-based inferred LRD luminosities at $z\approx5$.
We want to directly compare the upper-limits from \citet{Yue24} and \citet{Ananna24} (which are quoted in the rest-frame $2-10$ keV range) to LoRD X-ray luminosities. 
We use the X-ray SEDs of \citet{Revnivtsev18} and \citet{Balokovic18} (see also \citealt{Hickox18}) with varying column densities, $\log N_{H}=21, 22, 23 \text{ cm}^{-2}$ (i.e., the same regime explored by \citealt{Yue24}), and normalize the SEDs to the observed-frame $0.5-2$ keV luminosities of the LoRDs. 
We estimate their rest-frame $2-10$ keV luminosities by integrating the normalized X-ray SEDs over this energy range. 

\begin{table}[]
    \centering
    \begin{tabular}{lcc}
    \hline \hline
    XMM & Control & LoRDs \\
Band & $N=2166$ & $N=143$ \\
\hline
$0.5-2$ keV & 75.4$\pm$0.9\% & 77.6$\pm$3.5\%  \\ 
$0.5-10$ keV & 45.7$\pm$1.1\% & 49.0$\pm$4.2\% \\ 
$2-10$ keV & 30.0$\pm$1.0\% & 36.4$\pm$4.0\% \\ 
\hline
    \end{tabular}
    \caption{Detection percentages in the soft, full, and hard XMM-Newton bands for the control sample and the LoRDs. 
    The number, $N$, listed below the sample name is the number of sources which have an XMM source within 4\arcmin. 
    The uncertainty on the detection percentage is the standard deviation assuming a binomial distribution.}
    \label{tab:x-ray-properties}
\end{table}

The mean luminosities of the $\log N_H=\text{21, 22, 23 cm}^{-2}$ models for the X-ray detected LoRDs are $\log L_{2-10 \text{ keV}}=\text{43.4, 43.6, 44.4 erg s}^{-1}$, respectively. 
Given that broad line QSOs typically have minimal amounts of obscuration \citep[e.g.,][]{Koss17}, we focus our analysis on the $\log N_H=21 \text{ cm}^{-2}$ regime.
The rest-frame $2-10$ keV luminosity upper-limit of the AGN sample from \citet{Ananna24} is $\log L_{2-10 \text{ keV}}<43.28 \text{ erg s}^{-1}$; 44\% of all X-ray detected LoRDs fall below this upper limit in the low obscuration regime. 
Likewise, the mean upper-limit from \citet{Yue24} is $\log L_{2-10 \text{ keV}}<43.4 \text{ erg s}^{-1}$ and 54\% of LoRDs fall below this luminosity threshold.

\begin{figure*}
    \centering
    \includegraphics[width=\linewidth]{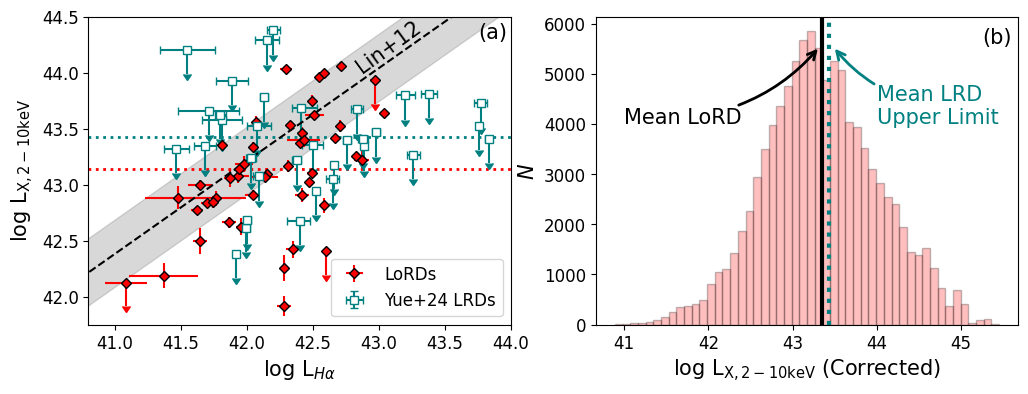}
    \caption{(a): Broad \ha luminosity versus rest-frame $2-10$ keV luminosity.
    Teal squares are LRD upper limits from \citet{Yue24} and red points are our sample of LoRDs, where diamonds with downward facing arrows indicate upper limits. 
    The black dashed line is the $L_X-L_{H\alpha}$ relation from \citet{Jin12} and the shaded region is the intrinsic scatter ($\sim0.3$ dex). 
    The dotted horizontal lines represent the mean of each sample. 
    (b): LoRD X-ray luminosity distribution after correcting for \ha luminosity (Equation \ref{eq:xray-scaled}). 
    The mean LoRD luminosity is the black solid line and the mean LRD upper-limit is the dashed teal line.
    LRD X-ray luminosity upper-limits are, on average, larger than the LoRD X-ray luminosities which may suggest that the average LoRD would go undetected in archival Chandra imaging at high-$z$.}
    \label{fig:Lx-LHa}
\end{figure*}

We show the $L_X-L_{H\alpha}$ relation, where $L_{H\alpha}$ is calculated from the broad component, for red2 LoRDs
and the \citet{Yue24} LRDs in Figure \ref{fig:Lx-LHa}~(a). 
The LoRDs largely sit at, or below, the \citet{Jin12} relation for broad line AGN, and the LRD upper limits are distributed both above and below the line; however, LRDs with large \ha luminosity ($\gtrsim10^{43} \text{ erg s}^{-1}$) are preferentially situated below the relation. 
We calculate the difference between the inferred $2-10$ keV luminosity and the expected luminosity from the \citet{Jin12} relation.
The average X-ray detected LoRD is situated $-0.25\pm0.49$ dex below the relation and the LRD upper limits are $-0.16\pm0.70$ dex below the relation, where the quoted uncertainty is the root-mean-square scatter. 
The control sample (not shown in Figure \ref{fig:Lx-LHa}~a) is $-0.20\pm0.45$ dex below the relation. 
The intrinsic scatter of the $L_X-L_{H\alpha}$ relation is $\sim0.3$ dex which suggests the LRD upper limits, the control sample, and the LoRDs are overall consistent with the relation. 

Differences in the optical luminosities may impact our interpretation as the LRD \ha luminosity distribution differs from our sample of control QSOs and LoRDs.
Therefore, we correct the $2-10$ keV X-ray luminosity distributions ($L_{X,\text{cor.}}$) according to Equation \ref{eq:xray-scaled}, 
\begin{equation} \label{eq:xray-scaled}
    L_{X,\text{cor.}} = L_{X} + \alpha (\log L_{\text{\ha,LRD}} - \log L_{\text{\ha}})
\end{equation}
where $\alpha$ is the slope of the $L_X-L_{H\alpha}$ relation ($\alpha=0.83$, \citealt{Jin12}).
We randomly sample the X-ray and \ha luminosities of our samples within their uncertainties and assign it to a \cite{Yue24} LRD source (sampled within the \ha uncertainty) and compute Equation \ref{eq:xray-scaled}. 
We perform 100,000 such iterations to estimate the scaled X-ray luminosity which is independent of \ha (Figure \ref{fig:Lx-LHa}~b). 
After correcting for the \ha luminosity, 57\% of LoRDs and 51\% of the control QSOs are below the \citet{Yue24} mean upper-limit. 


This is, of course, only applicable to the red2 LoRD sample, but we wish to determine how the full sample of LoRD $2-10$ keV luminosities (detections and upper limits) compare to the \citet{Yue24} upper limits. 
We treat the LoRD luminosity distribution as a left-censored, log-normal distribution. 
We determine the log-likelihood of the LoRD distribution by summing the probability density function of the detections ($N_{\rm{det}}=111$) and the cumulative density function of the upper limits ($N_{\rm{UL}}=18$); we maximize this to estimate the distribution's mean and standard deviation, i.e., $\log L_{X, \text{ LoRDs}}\sim\mathcal{N}(\mu=43.3, \sigma=0.50)$.
To propagate the uncertainty in the fit, we bootstrap resample (with replacement) the LoRD luminosities (detections and upper limits) 10,000 times, refit the distribution, and evaluate the cumulative distribution function at each LRD upper limit from \citet{Yue24}. 
The average probability that a randomly drawn LoRD is fainter than an LRD upper limit is $\bar{p}=0.57$ with a 95\% confidence interval (CI) of [0.52, 0.63]. 
If we split the LoRDs into V-LoRDs ($N_{\rm{det}}=29$, $N_{\rm{UL}}=4$) and non-V-LoRDs ($N_{\rm{det}}=28$, $N_{\rm{UL}}=4$) we find that the V-LoRDs are more likely to be less luminous than the LRDs ($\bar{p}=0.59$, CI [0.49, 0.69]) relative to the non-V-LoRDs ($\bar{p}=0.49$, CI [0.38, 0.60]).
We perform the same analysis on the control sample ($N_{\rm{det}}=1633$, $N_{\rm{UL}}=716$) and find the probability that a given control QSO is less luminous than an LRD upper limit is $\bar{p}=0.58$ (CI [0.57, 0.59]). 

Lastly, if we perform a Monte Carlo resampling analysis, where each low-$z$ luminosity is drawn from its uncertainty (upper limits are treated as a left-truncated normal distribution) and compared to an LRD upper limit, we find that $56\%$ of LoRDs and control QSOs are less luminous than the LRD upper limits.
Given that the average low-$z$ X-ray luminosity is less than the average LRD upper-limit, we conclude that it is not unreasonable to assume that a majority of the LoRDs (and control QSOs) would go undetected in archival Chandra observations at high-$z$. 
We discuss these findings, their implications, and the use of local X-ray scaling relations in Section \ref{subsec:xray-disc}.

\section{Discussion} \label{sec:disc-conc}


\subsection{Morphology} \label{subsec:morphology}
A key feature of high-$z$ LRDs is that the majority of sources appear unresolved in JWST imaging \citep[e.g.,][]{Greene24, Kokorev24, Kocevski25}; this inferred compact morphology also serves as an important argument for the AGN nature of LRDs. 
Recent work has shown that the rest-UV of LRDs is typically more extended than the rest-optical \citep{Cloonan26}. 
However, there exists significant uncertainty in modeling LRD sizes through traditional parametric fitting codes, especially in the low S/N regime \citep{Whalen26}.
We should note that classes of post starburst galaxies are known to be ultra-compact ($\lesssim100$ pc) as well \citep[e.g.,][]{Sell14, Whalen22, Davis23}.
The angular resolution of SDSS and JWST imaging greatly differ ($\sim1.2\arcsec$ compared to $<0.1\arcsec$), so selecting sources as ``unresolved" in SDSS imaging is entirely different than an unresolved source in JWST. 

Rather than selecting unresolved sources in SDSS, we use observations from the Hyper Suprime-Cam (HSC; \citealt{Aihara18}) survey which combines high angular resolution ($\sim0.4\arcsec$) and a wide survey area overlapping with SDSS. 
A $0.4\arcsec$ resolution corresponds to $\sim2.1$ kpc at $z=0.4$ and $\sim3.0$ kpc at $z=0.8$ which is still much larger than the inferred rest-optical LRD sizes ($R_{50} \lesssim100$ pc; \citealt{Cloonan26}); however, this is the highest resolution currently achievable in large‑scale ground‑based imaging.
We visually inspect $2\arcsec \times 2\arcsec$ $r$-band cutouts of LoRDs observed with HSC. 
$\sim23\%$ of LoRDs have suitable HSC imaging (61 V-LoRDs and 74 non-V-LoRDs) and $83\pm 2.2\%$ of those are unresolved, where the uncertainty is the binomial error.
$62\pm6.2 \%$ of V-LoRDs have an unresolved morphology compared to $89\pm3.6 \%$ of non-V-LoRDs. 
On average, resolved sources exhibit larger optical spectroscopic slopes ($1.0\pm0.7$, where $0.7$ is the standard deviation) relative to the unresolved sources ($0.35\pm1.0$), but the large uncertainties show significant overlap between the two samples. 

Given that resolved sources have a steeper optical slopes, we deduce that some of this flux must be coming from the stellar populations of the host galaxy. 
This is in agreement with our SED modeling (see Section \ref{subsec:seds} and Figures \ref{fig:seds} and \ref{fig:seds-lrds}) which requires a galaxy component to partly explain the optical flux. 
Interestingly, both the resolved and unresolved sources have the higher-order Balmer absorption features (i.e., H9-H12) and \caii H \& K in their stacked spectra (although \caii H and H$\epsilon$ are blended). 

In the resolved population the Balmer and \caii features can be straightforwardly interpreted as stellar atmospheric absorption.
In the unresolved population the situation is less clear; the detection of H9-H12 could be produced by dense, neutral hydrogen gas in the $n=2$ level along the line of sight. 
Such a population can be generated either by collisional excitation in very dense gas ($n\sim 10^{9-11} \text{ cm}^{-3}$; \citealt{Inayoshi25}) or by Ly$\alpha$ trapping \citep[e.g.,][]{Hall07, Aoki10}.
However, the simultaneous presence of \caii H \& K suggests a stellar component which can also produce the higher-order Balmer absorption. 
Disentangling these two scenarios (dense gas or stellar atmospheric absorption) would require modeling high-resolution, high-S/N spectral observations of these features on a source by source basis, or NIR spectroscopy of the Ca triplet \citep[e.g.,][]{Lin25}.
When considering the high-$z$ LRD spectra, Balmer absorption is primarily associated with the \ha and \hb emission lines, but the higher-order absorption features (i.e., H9-H12), if present, would likely be difficult to detect in LRD observations due to both small EWs and the low spectral resolution of PRISM observations (see Section \ref{subsec:spec} and Figure \ref{fig:prismified}a). 

\subsection{Spectral Comparison} \label{subsec:spec}

\begin{figure*}
    \centering
    \includegraphics[width=\linewidth]{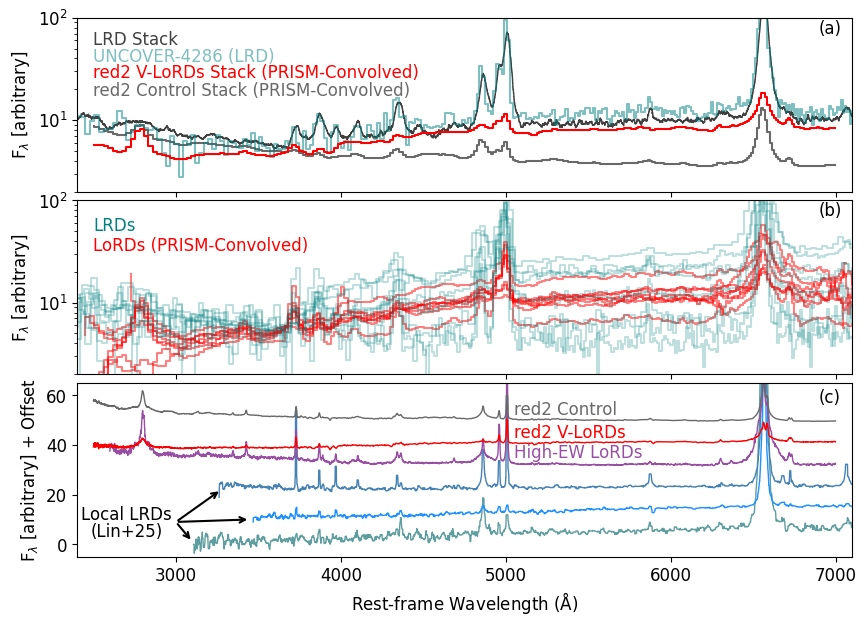}
    \caption{(a): JWST/PRISM spectrum for the LRD UNCOVER-4286 (teal) compared to a stacked SDSS spectrum of the red2 V-LoRDS (red) and the red2 control sample (gray). 
    The stacked SDSS spectra are convolved with the PRISM resolution and matched to the same wavelength grid as UNCOVER-4286. 
    A stack of LRDs from \citet{Graaff26} is shown in black.
    Each spectrum is normalized to the integrated flux at $3645\pm100\rm{\AA}$.
    (b): The same as panel~(a), except we show more individual LRDs (teal) and compare them to a sample of individual LoRDs (red); the LoRDs were selected to have spectral slopes similar to LRDs.
    (c): Spectral comparison between the local red dots of \citet{Lin25} (various shades of blue) and LoRD/control samples presented in this work. 
    We show the stacked red2 V-LoRD spectrum in red, a subset of high EW LoRDs in purple, and the red2 control sample in gray.}
    \label{fig:prismified}
\end{figure*}

The stacked LoRD spectra (Figure \ref{fig:stacked_spec}) do not look exactly like the ``typical" LRD observed with JWST/PRISM. 
These discrepancies are likely driven by four key factors: (1) the LoRDs are intentionally selected to be QSOs; (2) the low resolution of PRISM observations ($R\sim100$) only capture lines with the largest equivalent widths; (3) the stark Balmer break in some LRDs is not ubiquitous in our sample of LoRDs (although certain individual spectra do show this behavior); (4) our V-shape criteria differ in detail from those used in high-$z$ studies. 
We primarily use the V-LoRDs for illustrative purposes, so our different V-shape criteria does not bias our interpretation of the sample as a whole.

We intentionally select LoRDs from the DR16Q catalog to ensure the presence of broad Balmer lines, consistent spectral fitting, and secure redshifts \citep{Wu22}. 
However, this may bias our LoRD sample to QSO-like objects (i.e., sources with spectra similar to typical QSOs) and could alter our understanding of local LRD-like objects. 
Due to this, we clarify that our sample of LoRDs likely represents local LRD-like objects that have features consistent with typical QSOs.
This could explain why we do not find LoRDs with extreme Balmer breaks (although we note that other local-LRD searches also do not find these sources; e.g., \citealt{Lin26b}), and why the LoRDs have spectral properties generally consistent with QSOs (e.g., broad \mgii).
This may, in part, explain why a dominant AGN component is required when fitting the LoRD SEDs (Section \ref{subsec:seds}), but this does not impact the use of the AGN component when fitting the LRD SED (Figure \ref{fig:seds-lrds}b).
Applying the methodology introduced in this paper (i.e., filter matching in the rest-frame) to the full SDSS photometric catalog, then trimming the sample on broad Balmer emission may yield a more complete sample including unique sources we may have missed in our study.

At the same time, the stacked LoRD spectra visually differ from LRD spectra in part due to the dramatic difference in spectral resolution between JWST/PRISM and SDSS observations.
To illustrate the loss of spectral information in PRISM observations, we convolve the red2 V-LoRD stacked spectrum and the red2 control spectrum (Figure \ref{fig:stacked_spec}) with the wavelength-dependent PRISM dispersion and match their wavelength grids to an LRD.
In Figure \ref{fig:prismified}~(a) we show the PRISM-convolved SDSS spectra (red and gray) and compare these to UNCOVER-4286 (teal) and a stack of LRDs (black). 
We show UNCOVER-4286 \citep{Bezanson24} because it has a V-shaped continuum \citep{Setton24} and its redshift ($z=5.8$) means the spectral resolution is on the higher end. 
UNCOVER-4286 is representative of the median LRD in terms of its spectral shape and emission line properties.
The LRD stack shown in Figure \ref{fig:prismified} (a) is the median spectrum of the \citet{Graaff26} sources ($N=116$).

We show a sample of eight individual LoRD spectra relative to nine LRDs in Figure \ref{fig:prismified}~(b) to highlight both the similarity between LRDs and LoRDs as well as to illustrate the diversity of the LRD spectra. 
We select LoRD spectra from the red2 sample (to ensure coverage at longer wavelengths) which have $\beta_{\rm{opt}}>1.25$ and $\beta_{\rm{UV}}<0$, or $\beta_{\rm{opt}}-\beta_{\rm{UV}}>1.75$ \citep[c.f.,][]{Hviding25} for comparison purposes.
The LRD spectra in Figure~\ref{fig:prismified}~(b) are all unresolved V-shape sources \citep{Setton24}.
Specifically, we show JADES-68797/53501/13704/28074 \citep{Eisenstein26}, RUBIES-uds-40579/31747/47509 \citep{Graaff25a}, and UNCOVER-4286/45924 \citep{Bezanson24}. 

Only the brightest emission lines are clearly visible in the PRISM-convolved SDSS spectra: \mgii$\lambda$2800, \oii$\lambda$3727, H$\gamma$/\oiii$\lambda$4363, \hb, \oiii$\lambda$5007, and \ha. 
\mgii is not normally detected in individual LRD spectra, but this typical AGN line is weakly identified in stacks of LRD spectra \citep[e.g.,][]{Perez26}. 
The Balmer absorption features redward of \oii$\lambda$3727 are completely washed out and make \oii appear broader than it is.
Therefore, if LRDs have such features, we would be unable to detect them in existing PRISM observations.
This is consistent with the recent results which have found that absorption in the \ha emission line is detected at a higher rate when observed with higher spectral resolutions \citep{Matthee26}.
While this is not surprising, it highlights the importance of deep GRISM observations across the entire UV/optical spectrum where we may see typical stellar/AGN features too faint for PRISM level resolutions.

Similarly, the \nev$\lambda$3426 line, which is clearly visible in Figure \ref{fig:stacked_spec}, barely registers above the continuum in the PRISM-convolved spectrum (Figure \ref{fig:prismified}a). 
\citet{Labbe24} reports the detection of \nev$\lambda$3426 in the LRD A2744-45924 with deep PRISM observations; the presence of such high-ionization emission lines may be difficult to explain in the absence of a typical AGN SED. 
In the popular ``BH-star" model \citep[e.g.,][]{Begelman08, Inayoshi25, Naidu25}, the high-ionization emission lines are absorbed by a dense gas envelope surrounding the BH. 
Such emission lines are produced by the inner region of the accretion disk; if detected, this suggests a covering factor less than unity.
However, their presence means that even higher energy photons (i.e., X-rays) should also escape given that the photoionization cross section decreases with energy \citep[e.g.,][]{Inayoshi26}.
We explore this possibility in Section \ref{subsec:xray-disc}, and discuss the physical models in Section \ref{subsubsec:physical-models}.

The shape of the optical continuum in the LRD stack is similar to the red2 V-LoRD stack; both of which are clearly distinct from the control sample in the optical. 
The difference between the LoRD stack and the LRD stack in the UV is expected given that we do not enforce a negative UV slope when defining sources as V-shaped. 
The most pronounced difference between the LoRDs and the UNCOVER-4286 spectrum (or the LRD stack) is the EWs of the prominent emission lines (namely \ha and \hb), which is also apparent when we compare individual LoRDs and LRDs.  
LRDs, and high-$z$ AGN in general, are known to have large EWs -- roughly three times larger than local AGN \citep{Maiolino25}, but the nature of the large EWs in LRDs is uncertain. 
The differences between the LoRD and LRD EWs may be driven by variations in the covering factor, metallicity, ionization parameter, or host galaxy contamination. 
In star-forming galaxies the metallicity is known to decrease while the ionization parameter increases with increasing redshift \citep[e.g.,][]{Yuan13, Nakajima14,Cleri26}; this evolution is less understood in AGN partly due to selection effects \citep{Juneau14}, as well as difficulties in disentangling AGN from star-forming galaxies at high-$z$ \citep[e.g.,][]{Kewley13,Cleri25}. 

We compare our sample of LoRDs to the local red dots presented in \citet{Lin25}, one of which was independently discovered by \citet{Ji25b}. 
The three sources (SDSS J1047+0739, J1025+1402, and J1022+0841) were discovered in SDSS and reside at $z=0.1-0.2$; we show their SDSS spectra in Figure~\ref{fig:prismified}~(c) in different shades of blue. 
Similar to the LRDs, the \citet{Lin25} sources exhibit \ha and \hb EWs much larger than the red2 V-LoRD stack; to illustrate similarities between certain sources in our sample and the \citet{Lin25} sources we select a subset of LoRDs with large EWs. 
Specifically, we select LoRDs with \ha $\rm{EW} > 570\rm{\AA}$, or \hb $\rm{EW}> 110\rm{\AA}$; the \ha bound is set by photoionization models from \citet{Madau26} and the \hb bound is the mean EW of a stack of LRDs from \citet{Sun26}. 
There are 49 LoRDs which satisfy these criteria (\ha is redshifted out of the SDSS spectrographs for the red1 sample), we show the stacked spectrum for this sample in purple. 
In the high EW LoRD subset, we see more similarities in the emission line properties with the \citet{Lin25} sample compared to the full red2 V-LoRDs (e.g., H$\epsilon$, H$\delta$, H$\gamma$, \ion{He}{2}$\lambda$4686, \ion{He}{1}$\lambda$5876, \ion{Fe}{2}$\lambda$6369). 


\subsection{X-ray Detection Limits} \label{subsec:xray-disc}

LoRDs and the control sample are detected at similar rates in each X-ray band; these detection fractions are much larger than what we have come to expect for individual LRDs ($\sim0\%$).
However, as we showed in Section \ref{sec:x-ray}, the average LoRD X-ray luminosity (and control QSO luminosity) is at or below the average LRD X-ray upper limit of \citet{Yue24}. 
When we consider the shape of the LoRD optical continua, we find that V-shaped LoRDs are more likely ($\bar{p}=0.59$) to be less luminous than the LRD upper limits compared to non-V-shaped LoRDs ($\bar{p}=0.49$), but the total number of sources is small.  
Therefore, we cannot say for certain whether the LoRDs or typical QSOs would be detected in archival Chandra imaging at high-$z$. 

This highlights that the LRD X-ray upper limits are not all that constraining; therefore, the assumption that all LRDs will be undetected in archival (or new) Chandra imaging may not be a particularly sound assumption. 
Indeed the majority of LRDs are X-ray weak when considering well-established local relations.
The $L_X-L_{H\alpha}$ relation from \citet{Jin12} was constructed using a sample of 51 unobscured AGN with high quality optical and X-ray spectra at $0.03<z<0.38$ ($\langle z \rangle=0.14$).
They specifically exclude AGN with strong reddening effects and signatures of a warm X-ray absorber. 
Certain evidence suggests LRDs may have a non-negligible dust contribution \citep[e.g.,][]{Delvecchio25} which may cause the $L_X-L_{H\alpha}$ relation to fail. 
Given the observed differences between LRDs and the local AGN used to construct the $L_X-L_{H\alpha}$ relation (e.g., SED shape, Balmer line EWs, etc.), it is not unreasonable to think that LRDs may reside off this relation.
If this is the case, we cannot comment on their intrinsic X-ray weakness, only that LRDs (and LoRDs) do not abide by the same processes assumed in local type I AGN. 

\citet{Hviding26} reported an X-ray luminous ($L_{2-10 \text{ keV}}=10^{44.18} \text{ erg s}^{-1}$) source that has all the typical hallmarks of an LRD, which they refer to as the ``XRD" (see also ``The Forges," \citealt{Fu25}). 
The XRD is lower redshift ($z=3.1$) compared to the samples of \citet{Yue24} and \citet{Ananna24} ($z \approx 5.0$) and its X-ray luminosity is significantly larger than the LRD upper limits.
Depending on the dust attenuation, the XRD is either consistent with or below the $L_X-L_{H\alpha}$ relation and exhibits a moderate column density ($N_H\approx 2.4\times10^{22} \text{ cm}^{-2}$).
The authors suggest the XRD is an object transitioning between a gas enshrouded black hole and a typical AGN, where optically thin lines of sight towards the nucleus allow the X-rays to escape. 

When we consider that the LRD X-ray upper limits are poorly constraining, and that high-ionization emission lines are difficult to detect in current PRISM observations (Figure \ref{fig:prismified}a), this raises uncertainty regarding the popular ``BH-star" model. 
In this model, high energy accretion disk photons should be scattered/absorbed by the surrounding hydrogen envelope; detecting such photons implies a covering factor less than unity (i.e., there exist direct lines of sight between the accretion disk and gas beyond the broad line region), or a clumpy absorbing medium. 
Detections of such high-energy photons in LRDs have been observed \citep[e.g.,][]{Labbe24, Lambrides25, Tang25, Tripodi25, Ji26}, but the vast majority of X-ray observations have resulted in non-detections. 
This highlights the importance of deep, well-constraining X-ray observations of nearby LRD analog candidates, as well as the necessity of next generation X-ray observatories. 

\subsection{Implications for LRD Understanding} \label{subsec:implications}

\subsubsection{Informing Physical Models}
\label{subsubsec:physical-models}

In the ``BH-star" model of LRDs, dense, optically-thick gas surrounds the BH with a covering factor at, or very near, unity. 
The gas envelope absorbs and reprocesses the AGN continuum which is then reemitted as a (modified) blackbody \citep{Graaff26}; thus, the blackbody and AGN component cannot be simultaneously observed. 
This is in disagreement with our SED fits described in Section \ref{subsec:seds} (Figures \ref{fig:seds}d and \ref{fig:seds-lrds}b and c).
However, there are numerous explanations for the inclusion of the AGN and blackbody component. 
For instance, the non-spherical envelope model proposed by \citet{Lin25} contains optically-thin channels which allows AGN continuum photons to pass through the envelope.
Analyses of LRD narrow emission line ratios favor AGN excitation \citep[e.g.,][]{Sok26, Geris26} which implies that AGN-ionized photons must reach the interstellar medium (ISM). 
The detection of high-ionization emission lines in LRDs \citep[e.g.,][]{Labbe24, Tang25, Lambrides25} suggests AGN-ionization.
The frequent detection of broad and narrow Lyman-$\alpha$ emission in LRDs has been shown to be inconsistent with star-forming galaxies \citep[e.g.,][]{Tripodi25a, Tripodi25, Geris26, Morishita26} which further suggests that AGN-ionized photons reach the ISM.

Due to these lines of evidence, numerous studies favor a scenario wherein the dense gas envelope surrounding the SMBH is either clumpy with optically-thin lines of sight, or the covering factor is less than one \citep[e.g.,][]{Inayoshi25, Chen26, Ji26, Lambrides26, Lin25, Sok26}.
In such a configuration, a fraction of the AGN continuum escapes along the line of sight, and the rest is reprocessed in the optically-thick gas and then reemitted as a blackbody. 
\citet{Merida26} showed that such an SED is statistically favored over pure blackbody models when fitting LRD spectra; specifically, they find that 36/52 LRDs are best described by a hybrid model which includes stellar and/or AGN emission alongside blackbody emission. 
This is similar to our SED fits in Section \ref{subsec:seds}, but we emphasize that we remain agnostic to the physical process which produces the blackbody feature.

If the blackbody feature used in our SED fits originates from dense gas near the SMBH, then it would likely be subject to the same attenuation as the AGN component ($A_V=1.0$) unless a complicated geometry is invoked. 
Therefore, the intrinsic blackbody emission would be slightly hotter; we estimate this temperature by dereddening the observed blackbody using the \citet{Calzetti00} prescription and fitting a Planck function to the resulting curve. 
The temperature of the intrinsic blackbody is then $T\approx5800$K and $T\approx5000$K for the LoRDs (Figures \ref{fig:seds}d and \ref{fig:seds-lrds}c) and LRDs (Figure \ref{fig:seds-lrds}b), respectively. 

\subsubsection{Can LoRDs and LRDs be Understood as Reddened AGN?}
\label{subsubsec:connect-lrd-agn}

Our SED analysis (Section \ref{subsec:seds}) indicates that the typical emission from LoRDs is well-described by a reddened AGN combined with a young stellar population and a cool blackbody component. 
Therefore, the LoRDs appear to be a  subset of the larger reddened AGN population which exhibit broad lines (see \citealt{Montalvo26} for a broader discussion of these sources). 
This raises the question of what the LoRDs imply for the high-$z$ LRD population: namely, whether LRDs represent a distinct accretion mode from typical AGN and whether the LRDs form a homogeneous class of sources. 
Given that LoRDs share the same colors and certain spectroscopic characteristics as LRDs, a substantial fraction of LRD-selected sources may be understood as this subclass of reddened AGN. 
This is consistent with the statistical study of \citet{Billand26} which found that only a few percent of LRD-selected sources are physically distinct from the broader galaxy population, and the bulk of LRDs make up a continuous distribution of galaxies. 

The environments of LRDs provide independent support for the interpretation of these sources as a class of reddened AGN. 
A recent study by \citet{Barger26} found that the majority of LRD-selected sources in the Abell 2744 field have close companions ($<0.25$\arcsec). 
While close companions have important implications for the observed LRD properties, this result is also a natural expectation of the reddened AGN scenario: interactions and mergers are known to drive obscured nuclear activity. 
WISE-selected AGN, which include many obscured sources, are more likely to be found in interacting pairs than their unobscured counterparts \citep[e.g.,][]{Satyapal14, Weston17, Goulding18}, and the AGN merger fraction increases significantly with obscuration \citep[e.g.,][]{Glikman15, Gao20, Barrows23}. 
The prevalence of close companions among LRDs is therefore consistent with the interpretation of such sources as a class of reddened AGN. 
Whether this is also true for the LoRDs will be the subject of a future study.

On the other hand, attempting to unify LRDs as a single class of objects is tempting, but there is a clear diversity in spectral shapes among the LRD population \citep{Perez26}.
The LoRDs, if understood as a subclass of reddened AGN, are likely not representative of the most extreme LRD sources (e.g., ``The Cliff" or MoM-BH*-1, \citealt{Naidu25}). 
LRDs with the largest Balmer breaks are difficult to explain with reddened AGN models, as a steep extinction curve would be necessary to reproduce the spectral shape of the break \citep[e.g.,][]{Ma25, Graaff25}, though such sources make up the minority of what is currently referred to as an LRD \citep[e.g.,][]{Graaff25, Naidu25, Billand26, Perez26}. 
Therefore, distinguishing the exotic sources from sources that are consistent with a reddened AGN interpretation (at both high and low-$z$) are imperative for statistical studies of LRDs (e.g., the luminosity function, space densities, etc.) in order to inform their prevalence and evolution \citep[e.g.,][]{Matthee24, Kocevski25, Ma25, Ma25b, Greene26}.

{Deeper observations with JWST's Mid Infrared Instrument (MIRI; \citealt{Rieke15}) in future cycles will help to distinguish between reddened AGN and exotic sources in the LRD population. 
Current evidence of IR emission (presumably from dust) comes from stacked observations \citep[e.g.,][]{Delvecchio25, Perez26}, but if reddened AGN are included with the exotic sources, the dust emission may be underestimated.
The same argument applies to X-ray emission which can lead to an overestimation of the sample's X-ray weakness.
The origin of the narrow Balmer absorption in LRDs' Balmer lines is under debate and a complete sample would require deep medium/high-resolution spectroscopy \citep[e.g.,][]{Matthee26, Davis26}, but this feature is not seen in typical reddened AGN \citep{Shangguan26} and can be used to differentiate reddened AGN from the atypical sources (see the Appendix for LoRDs with potential Balmer absorption). 
Making such distinctions between sources is important for our understanding of the evolution of these enigmatic sources.

\section{Summary and Conclusions} \label{sec:conclusions}
In this paper we have shown that the JWST filters used to select LRDs at $z\sim5$ align well with the SDSS filters at both $z=0.38$ and $z=0.78$. 
We leverage this mapping to photometrically select Local Red Dots (LoRDs) from SDSS using selection criteria similar to the red1/red2 criteria from \citet{Greene24}; our selection is given in Equations \ref{eq:red1} and \ref{eq:red2}. 
In total we find $\sim1300$ QSOs which share similar photometric colors to the LRDs (Section \ref{subsec:selection}).
While not every LoRD we identify serves as a proper candidate analog to LRDs, this is the first step towards studying LRD-like systems in the local Universe at the population level where archival data is plentiful.

We fit the spectra using a broken power law in Section \ref{sec:spec} to select 244 LoRDs with V-shaped continua similar to those observed in LRDs.
The V-LoRDs exhibit a relatively flat UV continuum and a red rest optical. 
The stacked spectra show traditional AGN signatures such as \mgii, \nev, broad \ha and \hb, as well as higher-order Balmer absorption. 
The V-LoRDs have modest Balmer breaks similar to ``typical" LRDs \citep[c.f.,][]{Setton24}, but we do not identify any SDSS source with Balmer breaks as large as the most extreme LRDs \citep[c.f., ``The Cliff,"][]{Graaff25}.

A morphological analysis on a subset of LoRDs with HSC coverage shows that the majority of sources ($>80\%$) are compact regardless of their spectral shape (see Section \ref{subsec:morphology}). 
We find that LoRDs with a V-shaped continuum are more likely to have a resolved morphology compared to sources without the V-shape. 
This suggests that some of the V-shaped spectra ($\sim40\%$) are the result of an underlying stellar population and thus are unlikely to serve as proper LRD analogs. 
The remaining $\sim60\%$ of V-shaped LoRDs are proper LRD analog candidates, but space-based imaging is necessary to quantify their compactness.

Much of the spectral information in the stacked SDSS spectrum is lost when we convolve it with a PRISM-like resolution. 
This is important as information dense spectral features (e.g., high-ionization emission lines or stellar absorption features) may simply be undetectable in PRISM observations given the spectral resolution (see Section \ref{subsec:spec}). 
This highlights the importance of deep GRISM observations at all rest-UV/optical wavelengths. 
Likewise, higher-quality spectra on the LoRDs would help us to better understand the percentage of sources exhibiting narrow absorption in the broad Balmer emission lines (i.e., \ha and \hb, see the Appendix for LoRDs with potential absorption).

In Section \ref{subsec:seds}, we create composite SEDs of the control sample, LoRDs, and LRDs. 
The LRD and LoRD SEDs are consistent with each other in the rest-frame NUV to the NIR; both can be well modeled by a combination of a young stellar population (UV), a $T\sim4000$ K blackbody (optical) and a reddened AGN (optical/NIR). 
The LoRD SED is clearly distinct from the control QSO SED, particularly in the NUV/optical. 
The difference in the optical is due to an apparent Balmer break which can be modeled as either a cool blackbody or a starburst galaxy component. 
The reddening of the AGN component differentiates the LoRDs and control QSOs, where the LoRDs require more reddening.
Given the similarities between the LoRDs and LRDs, it is plausible that an important fraction of LRDs can be understood as a class of reddened AGN. 
Determining this fraction is essential for population level studies of these sources.

The LoRDs are detected at a similar rate compared to the control QSOs across each X-ray band. 
This detection fraction is much larger than what is expected from an LRD; however, we show that the LoRD (and control QSO) X-ray luminosities are consistent with the upper limits of LRD non-detections. 
We emphasize that the LRD X-ray upper limits are poorly constraining, regardless of whether or not the LoRDs are good analogs (see Sections \ref{sec:x-ray} and \ref{subsec:xray-disc}). 

Analogs of LRDs, or any high-$z$ galaxy for that matter, provide a useful testbed for theories and models of these enigmatic sources in a statistical sense. 
Galaxies in the nearby Universe have extensive archival multiwavelength coverage which are frequently of higher quality than the high-$z$ data and allow us to probe wavelength regimes which are inaccessible at high-$z$. 
Follow-up observations (e.g., deep X-ray exposures, space-based imaging, high signal-to-noise UV/optical spectroscopy) will be the true test as to what fraction of LoRDs are low-$z$ analogs of the LRDs. 
Such sources will enable us to garner a deeper understanding of this novel form of SMBH growth and galaxy evolution. 

\section{Acknowledgments}
We thank the anonymous reviewer, whose comments and suggestions greatly improved the content of this paper.
We thank Jenny Greene, David Setton, and Andy Goulding for helpful discussions. 
Q.O.C. and E.D. acknowledge support from the Dartmouth Fellowship. 
R.C.H. acknowledges support from NASA ADAP grant number 80NSSC25K7569.


Funding for the Sloan Digital Sky Survey IV has been provided by the Alfred P. Sloan Foundation, the U.S. Department of Energy Office of Science, and the Participating Institutions. 

SDSS-IV acknowledges support and resources from the Center for High Performance Computing  at the University of Utah. The SDSS website is www.sdss4.org.

SDSS-IV is managed by the 
Astrophysical Research Consortium 
for the Participating Institutions 
of the SDSS Collaboration including 
the Brazilian Participation Group, 
the Carnegie Institution for Science, 
Carnegie Mellon University, Center for 
Astrophysics | Harvard \& 
Smithsonian, the Chilean Participation 
Group, the French Participation Group, 
Instituto de Astrof\'isica de 
Canarias, The Johns Hopkins 
University, Kavli Institute for the 
Physics and Mathematics of the 
Universe (IPMU) / University of 
Tokyo, the Korean Participation Group, 
Lawrence Berkeley National Laboratory, 
Leibniz Institut f\"ur Astrophysik 
Potsdam (AIP),  Max-Planck-Institut 
f\"ur Astronomie (MPIA Heidelberg), 
Max-Planck-Institut f\"ur 
Astrophysik (MPA Garching), 
Max-Planck-Institut f\"ur 
Extraterrestrische Physik (MPE), 
National Astronomical Observatories of 
China, New Mexico State University, 
New York University, University of 
Notre Dame, Observat\'ario 
Nacional / MCTI, The Ohio State 
University, Pennsylvania State 
University, Shanghai 
Astronomical Observatory, United 
Kingdom Participation Group, 
Universidad Nacional Aut\'onoma 
de M\'exico, University of Arizona, 
University of Colorado Boulder, 
University of Oxford, University of 
Portsmouth, University of Utah, 
University of Virginia, University 
of Washington, University of 
Wisconsin, Vanderbilt University, 
and Yale University.

\hfil 

\textit{Software:} astropy \citep{astropy2022}, matplotlib \citep{matplotlib}, numpy \citep{numpy2020}, pandas \citep{pandas2021}, scipy \citep{scipy}, Topcat \citep{Taylor2005}.


\appendix 


\begin{figure}[h]
    \centering
    \includegraphics[width=0.5\textwidth]{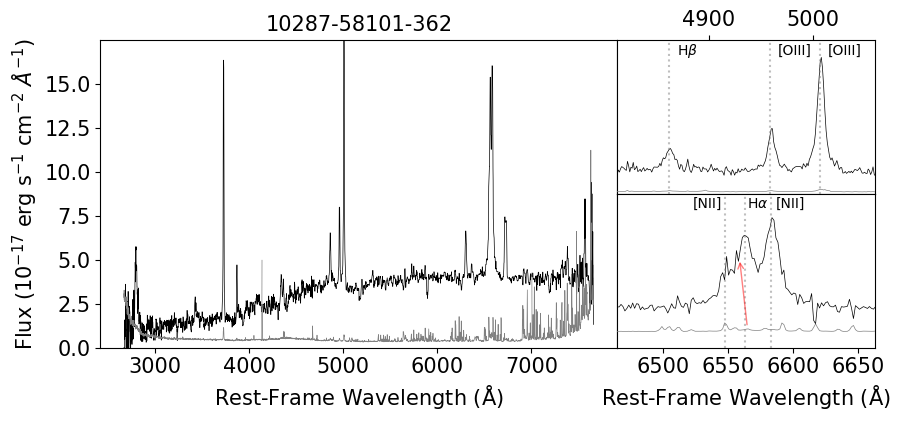}\hfill
    \includegraphics[width=0.5\textwidth]{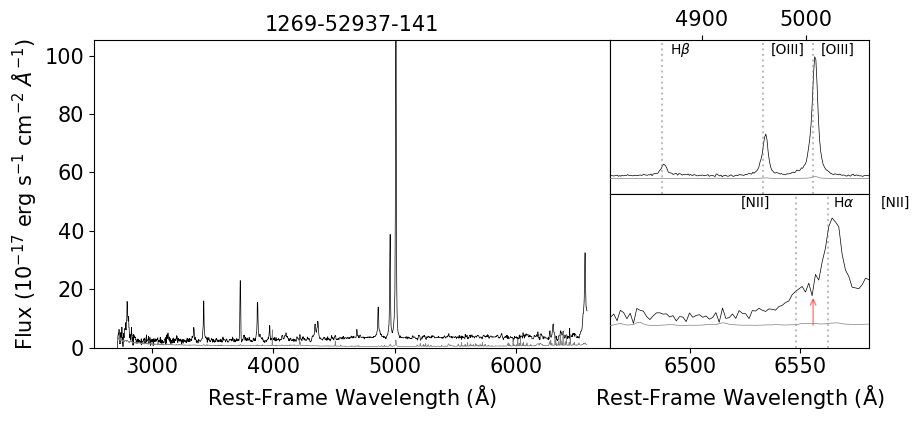}
    
    \includegraphics[width=0.5\textwidth]{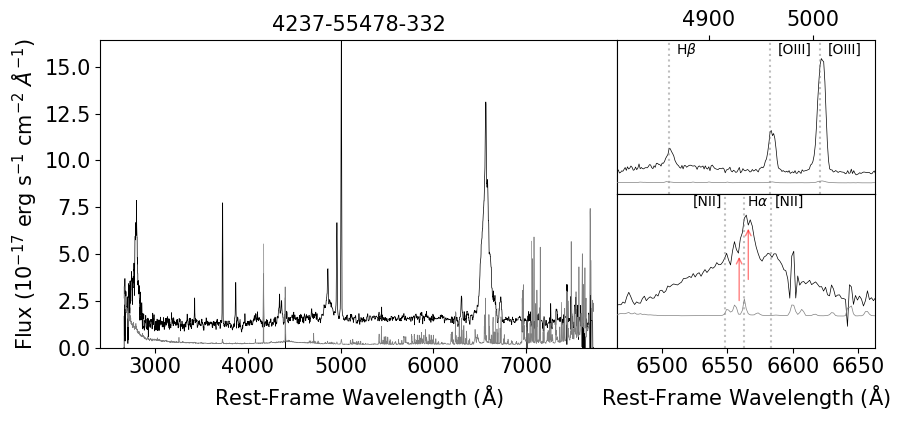}\hfill
    \includegraphics[width=0.5\textwidth]{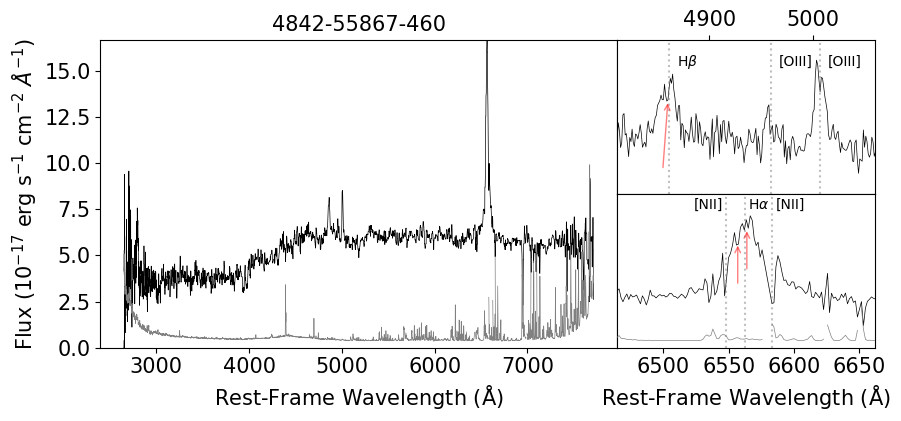}

    \includegraphics[width=0.5\textwidth]{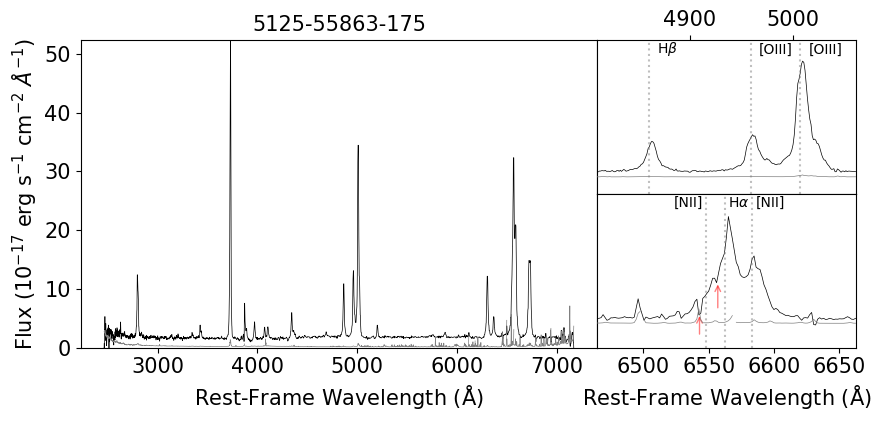}\hfill
    \includegraphics[width=0.5\textwidth]{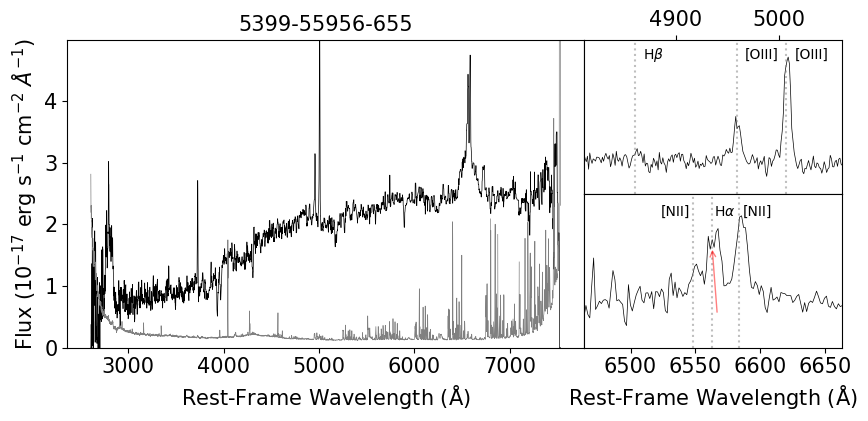}

    \includegraphics[width=0.5\textwidth]{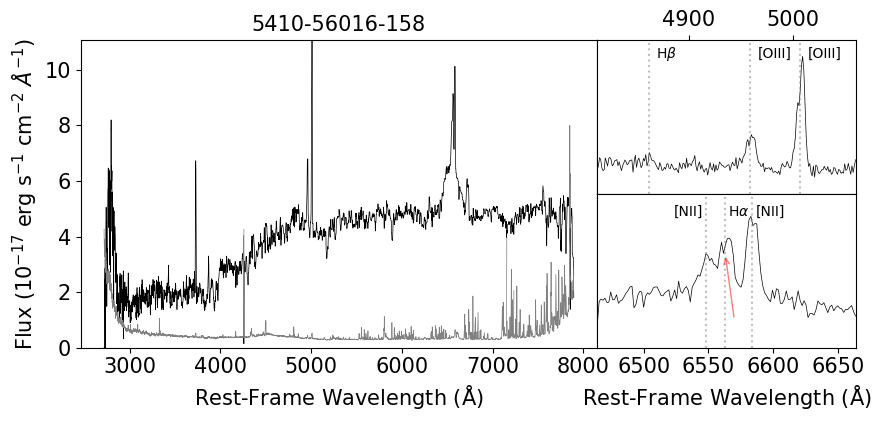}\hfill
    \includegraphics[width=0.5\textwidth]{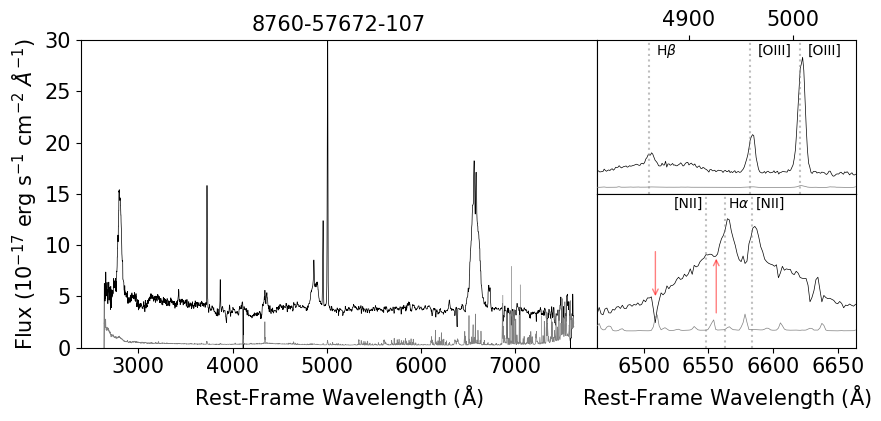}
    \caption{A subsample of the LoRDs which have signatures of absorption features in their Balmer lines (\ha and \hb).
    The left panel shows the full SDSS spectrum smoothed using SciPy's \cmtt{medfilt} function with a kernel size of seven. 
    The right panels zoom in on the \ha (bottom) and \hb (top) regions of the spectrum, and we do not apply any smoothing.
    Prominent emission lines are labeled, and potential absorption features are denoted with a red arrow. 
    The (unsmoothed) SDSS error spectrum is shown in gray.
    The title of the figure corresponds to the plate, MJD, and fiber of the SDSS observation.}
    \label{fig:appendix1}
\end{figure}

\begin{figure} \label{fig:appendix2}
    \centering
    \includegraphics[width=0.5\textwidth]{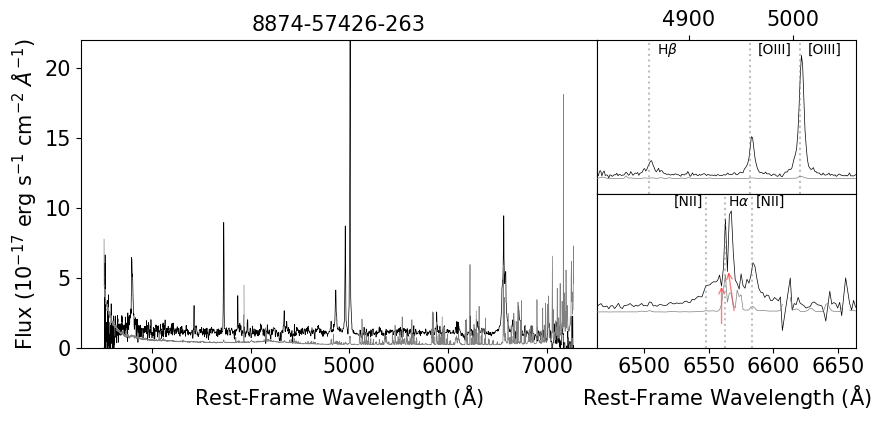}\hfill
    \includegraphics[width=0.5\textwidth]{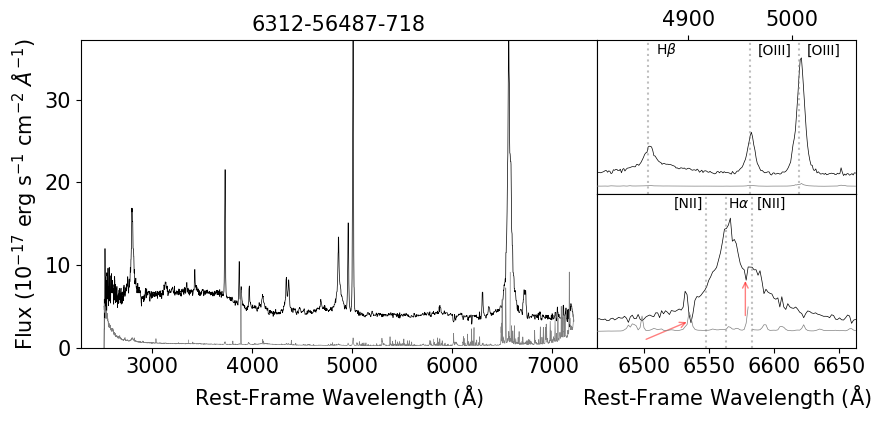}

    \includegraphics[width=0.5\textwidth]{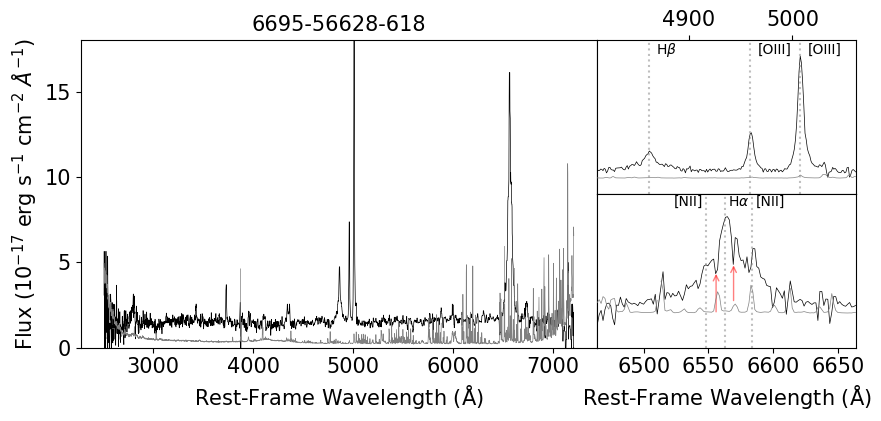}\hfill
    \includegraphics[width=0.5\textwidth]{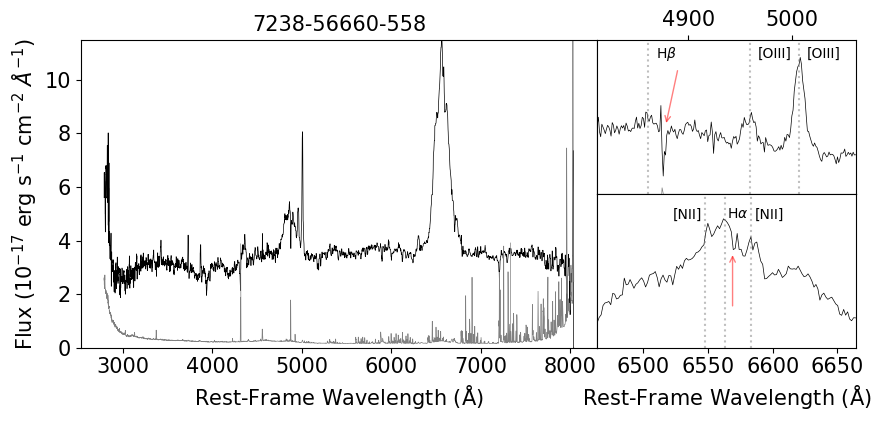}

    \includegraphics[width=0.5\textwidth]{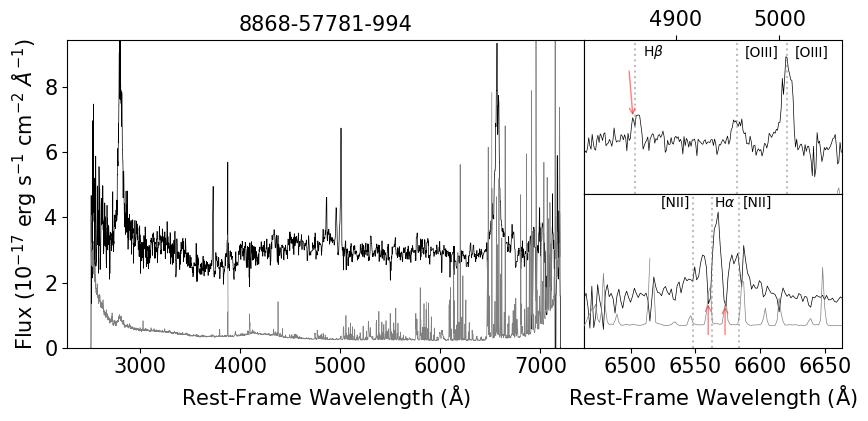}\hfill
    \includegraphics[width=0.5\textwidth]{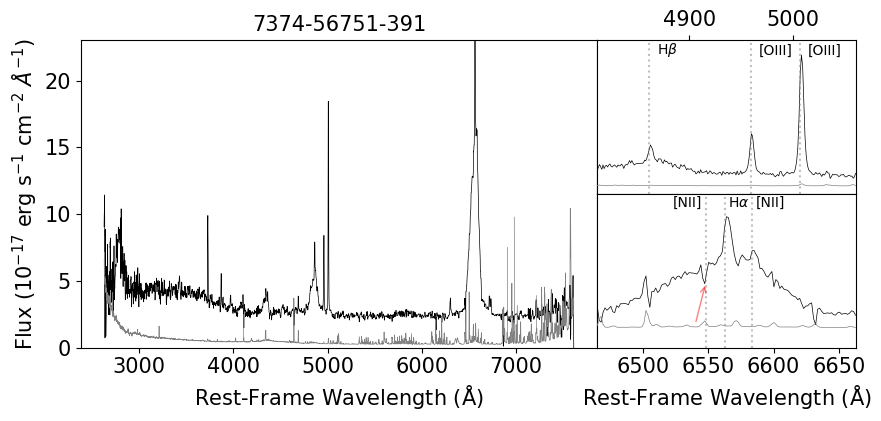}

    \includegraphics[width=0.5\textwidth]{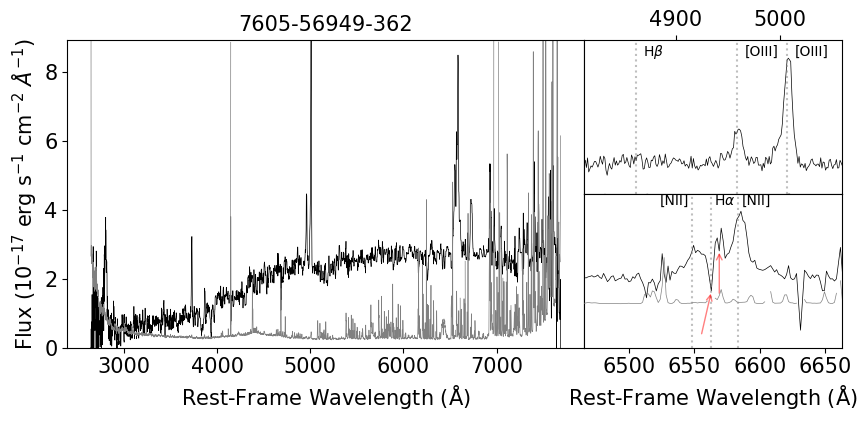}\hfill
    \includegraphics[width=0.5\textwidth]{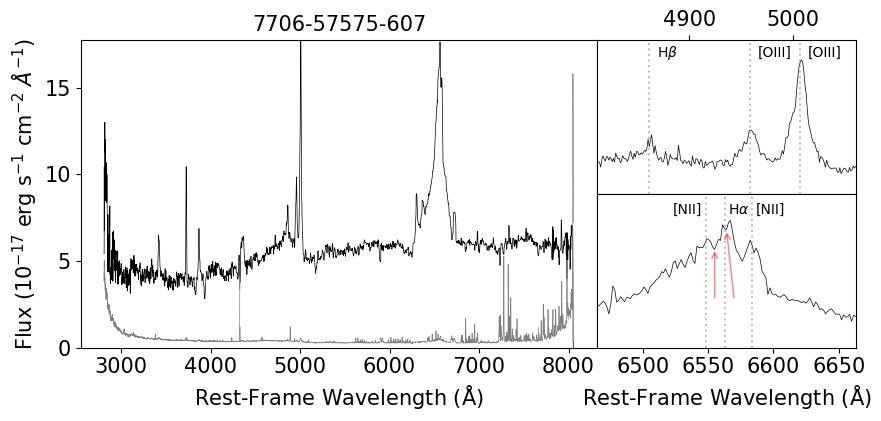}

    \includegraphics[width=0.5\textwidth]{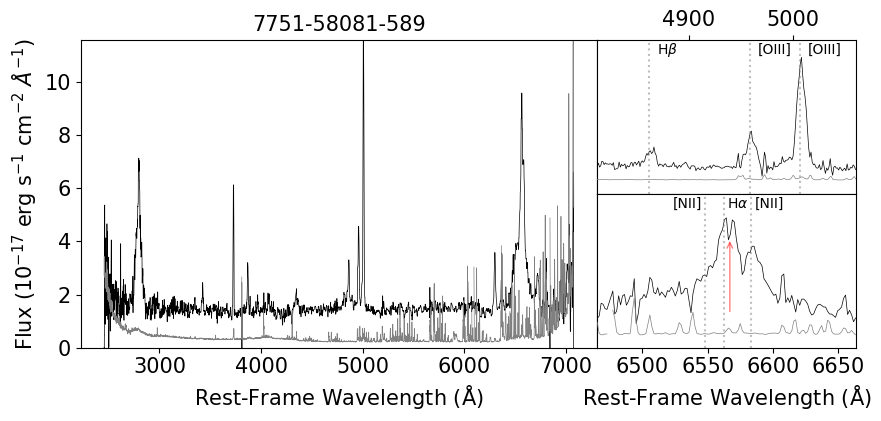}\hfill
    \includegraphics[width=0.5\textwidth]{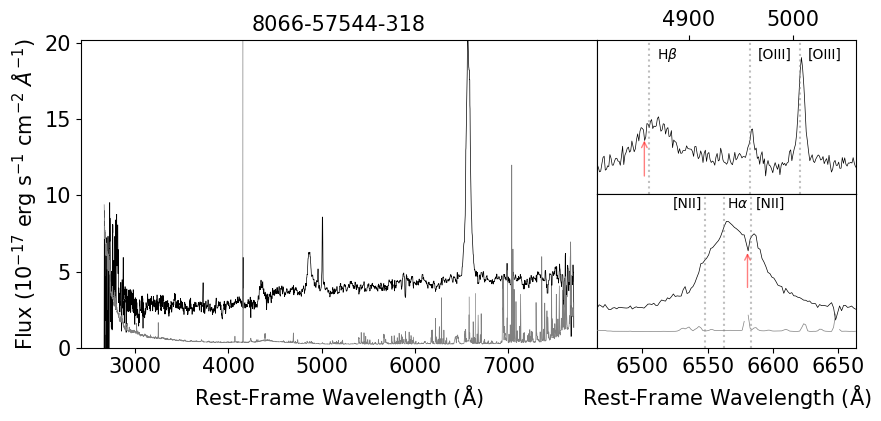}

    \caption{The same as Figure \ref{fig:appendix1}.}
\end{figure}

\end{document}